\documentclass[%
 reprint,
 amsmath,amssymb,
prb,
]{revtex4-2}
\usepackage{graphicx}
\usepackage{dcolumn}
\usepackage{bm}
\usepackage{amsfonts,braket}
\usepackage{hyperref}
\usepackage{color}
\usepackage[utf8]{inputenc}
\usepackage{csquotes}
\usepackage[T1]{fontenc}
\usepackage{multirow, tabularx}
\usepackage{siunitx, booktabs}
\usepackage{relsize}
\newcolumntype{Z}{D..{4.2}}
\newcolumntype{Y}{>{\centering\arraybackslash}X}
\newcolumntype{C}[1]{>{\centering\arraybackslash}p{#1}}
\graphicspath{{figures/}}
\usepackage[section]{placeins}

\newcolumntype{d}{D{.}{.}{-1}} 

\newcommand{\qpoint}{\mbox{$\mathbf{q}$-point} }
\newcommand{\kpoint}{\mbox{$\mathbf{k}$-point} }
\newcommand{\subkn}[2]{_{\mathbf{#1}{#2}}}
\newcommand{\kn}{_{\mathbf{k}{n}}}
\newcommand{\qv}{_{\mathbf{q}{\nu}}}
\newcommand{\wqv}{\omega_{\mathbf{q}{\nu}}}
\newcommand{\gru}{Gr\"uneisen }
\DeclareMathOperator{\e}{e}

\begin{document}


\title{
Zero-point lattice expansion and band gap renormalization: \gru approach versus free energy minimization}

\author{V\'eronique Brousseau-Couture}
\email{veronique.brousseau.couture@umontreal.ca}
\affiliation{%
D\'epartement de physique, Universit\'e de Montr\'eal, C.P. 6128, Succursale Centre-Ville, Montr\'eal, Qu\'ebec, Canada H3C 3J7
}%

\author{\'Emile Godbout}%
\affiliation{%
D\'epartement de physique, Universit\'e de Montr\'eal, C.P. 6128, Succursale Centre-Ville, Montr\'eal, Qu\'ebec, Canada H3C 3J7
}%
\author{Xavier Gonze}%
\affiliation{%
Institute of Condensed Matter and Nanosciences, UCLouvain, B-1348 Louvain-la-Neuve, Belgium
}%
\author{Michel C\^ot\'e}%
\affiliation{%
D\'epartement de physique, Universit\'e de Montr\'eal, C.P. 6128, Succursale Centre-Ville, Montr\'eal, Qu\'ebec, Canada H3C 3J7
}%

\date{\today}

\begin{abstract}
The zero-point lattice expansion (ZPLE) is a small variation of the lattice parameters induced by the presence of phonons in a material compared to the static lattice picture. 
It contributes significantly to the zero-point renormalization (ZPR) of the band gap energy, but its consequences have not been investigated as thoroughly as those stemming from electron-phonon interactions. 
In the usual first-principles approach, one evaluates the ZPLE by minimizing the $T=0$~K Helmholtz free energy. 
In this work, we show that the formalism based on the \gru parameters, which commonly neglects zero-point effects, can be efficiently used to compute ZPLE for both isotropic and anisotropic materials at much lower computational cost. 
We systematically test this formalism on 22 cubic and wurtzite materials and obtain excellent agreement with free energy minimization results for both the ZPLE and the resulting band gap ZPR. 
We use our results to validate an empirical expression estimating the ZPLE-induced ZPR and unveil its sensitivity to the temperature range involved in estimating the ZPLE from experimental data. 
Our findings finally reveal that the ZPLE contribution to the band gap ZPR can reach 20\% to more than 80\% of the electron-phonon interaction contribution for heavier or more ionic materials, including materials containing light atoms. 
Considering both contributions on an equal footing is thus essential should one attempt to compare theoretical ZPR results with experimental data. \\
\newline
\textit{This is a post-peer-review version of the article accepted for publication in Physical Review B, which includes the Supplemental Material in the main file.}
  
\end{abstract}

\maketitle

\section{\label{sec:intro}Introduction}

The electronic band gap and the lattice parameters are undoubtedly two of the most fundamental properties of materials. The former depends on the latter, as a modification of the lattice parameters modifies the ionic potential felt by the electrons, hence affecting the electronic eigenvalues. This is at the heart of band gap engineering through lattice mismatch~\cite{kuo_effect_1985}. While the standard Kohn-Sham DFT is not theoretically expected to deliver accurate excited-state properties such as band gap energies~\cite{martin_electronic_2004}, it should in principle correctly predict the lattice parameters, as they are a ground state property.

The standard DFT framework assumes, however, a perfectly static lattice. This picture would be physically correct should the ions behave like classical particles as $T\rightarrow0$~K. This is not the case in the quantum mechanical picture: quantum fluctuations of the nuclei positions remain present even at absolute zero temperature. This phenomenon, known as the zero-point (ZP) motion of the ions, is responsible for the zero-point renormalization (ZPR) of the band gap. The leading mechanism driving the ZPR is electron-phonon interaction (EPI), namely, the electronic response to the distorted ionic potential induced by the lattice vibrations. In addition, the phonon zero-point energy contributes to the total free energy of the crystal, which, in turn, affects the equilibrium lattice parameter at $T=0$~K compared to the static value. This lattice parameter shift, known as the zero-point lattice expansion (ZPLE), also affects the band gap energy, thus contributing to the ZPR.

EPI has been widely investigated by the first-principles community throughout the last two decades~\cite{giustino_electron-phonon_2017}. In addition to the ZPR and the temperature dependence of the electronic structure, it is at the core of transport properties~\cite{ponce_first-principles_2020, brunin_phonon-limited_2020}, polaronic systems~\cite{verdi_origin_2017,nery_quasiparticles_2018} and superconductivity~\cite{giustino_electron-phonon_2007, blase_role_2004}. 
ZPLE has been the subject of less attention by this community. 
It has been mostly considered in the broader context of thermal expansion and temperature-dependent thermodynamical quantities rather than as a zero-point correction itself~\cite{huang_efficient_2016, lejaeghere_ab_2014, erba_combining_2014, otero-de-la-roza_treatment_2011, mounet_first-principles_2005, rignanese_ab_1996}. It has also been addressed when investigating the accuracy of lattice parameters predictions by different exchange-correlation functionals~\cite{hao_lattice_2012,csonka_assessing_2009, haas_calculation_2009, schimka_improved_2011}, as this correction is essential to make a fair comparison to experimental values. In all cases, most approaches rely on minimizing the Helmholtz free energy (FE) with respect to volume. Some works have also proposed a more generalized framework which allows for treatment of anisotropic thermal expansion~\cite{lichtenstein_anisotropic_2000,carrier_first-principles_2007, rasander_thermal_2018,pike_calculation_2019}.

Another method, developed by \gru in the early 1900's~\cite{gruneisen_theorie_1912}, expresses the temperature-dependent thermal expansion coefficient in terms of the volume variation of the phonon frequencies. The main advantage of this method is its smaller numerical cost, as it is based on finite derivatives rather than the computation of the full temperature and volume-dependent free energy. However, in its typical form, the \gru formalism neglects ZPLE, most likely for historical reasons, as this formalism was originally intended to model the thermal expansion coefficient, which does not depend on the zero-point energy. Nevertheless, some authors recognized that, in principle, the ZPLE could be properly incorporated within this formalism~\cite{huang_efficient_2016, allen_quasi-harmonic_2019}. To our the best of our knowledge, no systematic study of the predictive capabilities of the \gru formalism compared to the standard FE minimization technique regarding ZPLE has been attempted.

Despite being a decades-old problem, developing a first-principles-based methodology that can accurately predict thermal-dependent material properties in a wide range of temperatures is still an active subject of investigation. Recent proposals include a fully anharmonic one-shot formalism that can account simultaneously for both EPI and thermal expansion~\cite{zacharias_fully_2020}, formalisms for temperature-dependent elastic constants~\cite{rignanese_ab_1996,wu_quasiharmonic_2011,malica_quasi-harmonic_2020}, self-consistent volume change calculations~\cite{huang_efficient_2016}, incorporation of higher-order phonon-phonon interactions within the \gru formalism~\cite{masuki_anharmonic_2021} and accounting for fully anharmonic temperature-dependent phonon frequencies~\cite{pike_calculation_2019, hellman_temperature_2013}. All these methodologies could, in principle, be used to assess the ZPLE.  However, we can reasonably expect that the additional true anharmonicity captured by these methodologies will not significantly affect the $T=0$~K ZPLE for weakly or moderately anharmonic materials.

On the other hand, the contribution of thermal expansion to the temperature-dependent band gap energy has been investigated in semiconductors~\cite{querales-flores_temperature_2019, monserrat_temperature_2016,bravic_finite_2019, zhang_temperature-dependent_2020, wu_first-principles_2018}, layered materials~\cite{villegas_anomalous_2016, monserrat_temperature_2017}, nanostructures~\cite{kamisaka_ab_2008,abdulsattar_effect_2019} and topological materials~\cite{querales-flores_towards_2020, brousseau-couture_temperature_2020, monserrat_unraveling_2019}, mostly on a case-by-case basis. Outside this broader context, the explicit study of the importance of the ZPLE contribution to the $T=0$~K band gap renormalization compared to the one stemming from EPI is more scarcely addressed throughout the literature, a noticeable exception being the work of Garro \textit{et al.}~\cite{garro_dependence_1996}. Furthermore, it is typically thought of as negligible compared to the EPI contribution, without tangible statistical evidence. General trends allowing to qualitatively estimate the importance of ZPLE on the band gap ZPR in terms of simple bulk material properties are also still lacking, even for benchmark materials like diamond and zincblende structures.

In this work, we take a new look at the ZPLE and its contribution to the band gap ZPR from the first-principles perspective.  We investigate 22 materials: 19 with cubic symmetry and 3 with wurtzite structure. We show that the well-established \gru formalism can account for the ZPLE and its effect on the band gap as accurately as the typical FE minimization approach, for both isotropic and anisotropic materials, at a much lower computational cost. We also validate an empirical expression that estimates the ZPLE band gap correction from simple parameters for sufficiently symmetric materials and discuss its possible shortcomings when used in conjunction with experimental data. We finally demonstrate that, for most materials in our set, the ZPLE modifies the predicted band gap ZPR from EPI by more than 20\%, even for materials containing light atoms, thus questioning the perception that ZPLE effects can be neglected in such materials.

The remainder of this paper is organized as follows. In Sec.~\ref{sec:theory}, we briefly review the theoretical formalism used to compute the ZPLE and the EPI contributions to the band gap ZPR and demonstrate how to evaluate the ZPLE within the \gru formalism. The relevant computational details are summarized in Sec.~\ref{sec:computation}. Sections~\ref{sec:res-ZPLE} and \ref{sec:res-compare} respectively present our numerical results for the ZPLE and compare the free energy minimization and \gru approaches, while Sec.~\ref{sec:res-gap} focuses on the ZPLE contribution to the band gap renormalization. We finally discuss some implications of our results in Sec.~\ref{sec:discussion} and summarize our findings.
\section{\label{sec:theory}Methodology}
\subsection{Zero-point renormalization of the band gap energy}\label{sec:zpcorr}
Electronic eigenstates depend both on the unit cell structure and temperature. For a general material, the unit cell structure is fully defined by the lattice parameters, $\{a_i\}$, the angles between them, $\{b_i\}$, and the internal atomic coordinates, $\{z_i\}$. For lattice parameters and angles, the index $i=1,2,3$ refers to the three independent lattice vectors, while for the internal atomic coordinates, $i$ runs over the different atoms in the unit cell. We introduce the condensed notation  
\begin{equation}
   \{V\} \triangleq \{\{a_i\}, \{b_i\}, \{z_i\}\},
\end{equation}
to denote the full set of structure parameters.

We wish to evaluate the temperature dependence of an electronic eigenvalue $\varepsilon\kn$, where $\mathbf{k}$ is the electron wavevector and $n$ is the branch index. From thermodynamic theory, we can express the variation of $\varepsilon\kn$ with respect to temperature at constant stress $\bm{\sigma}$ (typically, at constant pressure) in terms of its variation at a fixed cell structure and of the structural variation at constant stress and temperature:
\begin{equation}\label{eq:thermo-general}
    \begin{split}
    \left(\frac{\partial\varepsilon\kn(\{V\},T)}{\partial T}\right)_{\bm{\sigma}} = &\left(\frac{\partial\varepsilon\kn(\{V\},T)}{\partial T}\right)_{\{V^0\}}\\ + &\left(\frac{\partial\varepsilon\kn(\{V\},T)}{\partial \{V\}}\right)_T\left(\frac{\partial\{V\}}{\partial T}\right)_{\bm{\sigma}}.
    \end{split}
\end{equation}
The first term describes the effect of the electron interaction with a given phonon population at constant equilibrium configuration $\{V^0\}\triangleq \{\{a_i^0\}, \{b_i^0\}, \{z_i^0\}\}$, which minimizes total electronic energy in the static lattice approximation, that is, the EPI contribution that will be defined in Sec.~\ref{sec:theory-epi}. The second term captures the effect of the thermal modification of the lattice, including all variations of the lattice parameters, lattice angles and internal coordinates, and can be seen as a deformation potential of the band gap~\cite{cardona_isotope_2005}. 

Working in a quasiparticle framework (e.g. DFT or GW approximations), one can obtain the  temperature-dependent eigenvalue correction to the eigenvalue for static lattice, $\varepsilon^0\kn$, by integrating Eq.~(\ref{eq:thermo-general}):

\begin{equation}\label{eq:eigcorr}
\begin{split}
        \Delta\varepsilon\kn (T) &= \varepsilon\kn (T) - \varepsilon^0\kn (\{V^0\})\\
    &\approx \int\limits_0^T dT' \left.\frac{\partial\varepsilon\kn(\{V\},T')}{\partial T'}\right|_{\{V^0\}} \!\!+ \textrm{ZPR}\kn^{\textrm{EPI}}\\
    &+ \int\limits_0^{T}dT' \left.\frac{\partial\varepsilon\kn(\{V\},T')}{\partial \{V\}}\right|_{T'}\!\!\left.\frac{\partial\{V\}}{\partial T'}\right|_{\bm{\sigma}} \!\!+ \textrm{ZPR}\kn^{\textrm{ZPLE}}\\
    &=\Delta\varepsilon\kn^{\rm{EPI}}(T) + \textrm{ZPR}\kn^{\textrm{EPI}} + \Delta\varepsilon\kn^{\rm{TE}}(T) + \textrm{ZPR}\kn^{\textrm{ZPLE}},
    \end{split}
\end{equation}
where $\Delta\varepsilon\kn^{\rm{EPI}}$ and $\Delta\varepsilon\kn^{\rm{TE}}$ respectively stand for the electron-phonon interaction and thermal expansion (TE) contribution to the renormalization. Note that the static equilibrium eigenvalues are implicitly computed at \mbox{$T=0$~K}. The lower bound of both integrals refers to the \mbox{\textit{dynamical}} $T=0$~K, which considers both electrons and ions. The two integration constants, $\textrm{ZPR}\kn^{\textrm{EPI}}$ and $\textrm{ZPR}\kn^{\textrm{ZPLE}}$, capture the effect of the zero-point motion of the ions on the $T=0$~K modification of the static electronic eigenenergies. We thus implicitly suppose that the true ZPR$\kn$ correction, which corresponds to the EPI interaction computed at the $T=0$~K renormalized lattice structure, can be evaluated in two distinct steps, hence the similar symbol on the second line. The band gap ZPR, which we define using the symbol ZPR$_g$, can thus be approximated at first order by the sum of the EPI and 
lattice deformation contributions~\cite{allen_temperature_1983}:
\begin{equation}\label{eq:sumzpr-full}
    \textrm{ZPR}_g \simeq \textrm{ZPR}_g^{\textrm{EPI}} + \textrm{ZPR}_g^{\textrm{ZPLE}}. 
\end{equation}

We now focus on the contribution of lattice deformation. We can recast the second term of Eq.~(\ref{eq:thermo-general}) by making explicit the derivative with respect to $\{V\}$ in terms of derivatives with respect to the lattice parameters, lattice angles and internal coordinates:
\begin{equation}\label{eq:derivativesfullsum}
\begin{split}
    &\left(\frac{\partial\varepsilon\kn(\{V\},T)}{\partial \{V\}}\right)_T\left(\frac{\partial\{V\}}{\partial T}\right)_{\{\sigma_i\}}\\ 
    &= \sum\limits_{j=1}^3 \left(\frac{\partial\varepsilon\kn(\{V\}, T)}{\partial a_j}\right)_{\substack{\{a_{i\neq j}\},\\ 
    \{b_i\}, \{z_i\}}}
    \left(\frac{\partial a_j}{\partial T}\right)_{\{\sigma_i\}} \\
    &+ \sum\limits_{j=1}^3 \left(\frac{\partial\varepsilon\kn(\{V\}, T)}{\partial b_j}\right)_{\substack{\{b_{i\neq j}\},\\ 
    \{a_i\}, \{z_i\}}}
    \left(\frac{\partial b_j}{\partial T}\right)_{\{\sigma_i\}} \\
    &+ \sum\limits_{j} \left(\frac{\partial\varepsilon\kn(\{V\}, T)}{\partial z_j}\right)_{\substack{\{z_{i\neq j}\},\\ 
    \{a_i\}, \{b_i\}}}
    \left(\frac{\partial z_j}{\partial T}\right)_{\{\sigma_i\}}.
    \end{split}
\end{equation}
In this expression, all the structural parameters kept constant are set at their static equilibrium value, $\{V^0\}$.

We can, however, circumvent the explicit evaluation of the derivatives contained in Eq.~(\ref{eq:derivativesfullsum}) by supposing that the temperature-dependent eigenvalue correction induced by lattice deformation ($\Delta\varepsilon\kn^{\textrm{TE}}$, third line of Eq.~(\ref{eq:eigcorr})) can be approximated from the static DFT eigenvalue evaluated at the temperature-dependent lattice structure, $\{V(T)\}$. As will be made explicit in the following section (Sec.~\ref{sec:theory-ZPLE}), we furthermore suppose that the leading contribution to $\Delta\varepsilon\kn^{\textrm{TE}}$ comes from the variation of the lattice parameters and lattice angles, and work with the set of structural parameters $\{\tilde{V}(T)\}$, in which the internal coordinates are chosen to minimize the static DFT total energy at fixed lattice vectors. We thus obtain
\begin{equation}\label{eq:eigcorr-effective}
    \Delta\varepsilon\kn^{\textrm{TE}}(T) \approx \varepsilon\kn^{\textrm{DFT}}(\{\tilde{V}(T)\}) - \varepsilon\kn^{\textrm{DFT}}(\{V^0\}).
\end{equation}
Within these approximations, $\textrm{ZPR}_g^{\textrm{ZPLE}}$ becomes the difference between the DFT band gap energy evaluated at the $T=0$ dynamical lattice structure including zero-point motion effects and the one evaluated at the static equilibrium structure.

\subsection{Zero-point lattice expansion in the quasiharmonic approximation}\label{sec:theory-ZPLE}
\subsubsection{Free energy formalism}\label{sec:theo-fe}
Throughout this work, we use the Hartree atomic unit system \mbox{($\hbar=m_e=c=|e|=1$)}. At constant external stress~$\{\bm{\sigma}\}$, the Helmholtz free energy (FE) of the crystal can be approximated by the sum of the electronic and vibrational contributions:
\begin{equation}\label{eq:ftot-helm-full}
\begin{split}
F(\{V\}, T) &= F^{\textrm{e}}(\{V\})+F^{\textrm{vib}}(\{V\}, T)\\
&= E_{\textrm{stat}}^{\textrm{e}}(\{V\}) - k_B T \ln Z^{\textrm{ph}}(\{V\}, T),
\end{split}
\end{equation}
where the electronic part is further approximated by the total energy obtained at a given structural configuration $\{V\}$ in the static lattice approximation, since we are dealing with semiconductors and insulators, whose entropic contribution to the FE can be safely neglected. $k_B$ is the Boltzmann constant.

From the phonon partition function in the harmonic approximation, $Z^{\textrm{ph}}$, the vibrational free energy can be decomposed into the zero-point (ZP) and thermal (th) contributions: 
\begin{equation}\label{eq:fvib-vol}
    \begin{split}
        F^{\text{vib}} &= -k_B T \ln (Z^{\text{ZP}} Z^{\text{th}})\\
        &= \sum\limits\qv \frac{\wqv(\{V\})}{2} + k_B T \sum\limits\qv \ln\left(1-\e^{-\frac{\wqv(\{V\})}{k_B T}}\right)\\
        &\triangleq F^{\text{ZP}}(\{V\}) + F^{\text{th}}(\{V\},T),
    \end{split}
\end{equation}
where the contributions of all phonon modes with wavevector $\mathbf{q}$, branch index $\nu$ and frequency $\wqv$ are summed for all branch indices and all wavevectors in the Brillouin zone (BZ).
The first term of this expression is the well-known zero-point energy, which is responsible for the ZPLE. In this equation and for the remainder of this work, all summations over the phonon modes are implicitly normalized by the number of wavevectors in the BZ sampling.

In the quasiharmonic approximation (QHA), the main contribution to the temperature dependence of the phonon frequencies is represented by their variation with respect to the lattice parameters, which has the effect of causing thermal expansion at finite temperature and ZPLE at $T=0$~K. It should be understood that the purpose of the QHA is not to capture the full anharmonicity of the interatomic force constants at a fixed structure, which is the true physical origin of thermal expansion and ZPLE, as would do non-perturbative methods or the TDEP method~\cite{hellman_temperature_2013}. It rather provides us with an accurate leading-order expression for the temperature-dependent 
structural change of the crystal~\cite{allen_quasi-harmonic_2019}. For most materials, the true anharmonicity correction to the ZPLE beyond QHA can be reasonably expected to be negligible, as we investigate weakly anharmonic materials and restrict ourselves to $T=0$~K (see Ref.~\cite{allen_quasi-harmonic_2019}).

The temperature-dependent lattice, $\{V(T)\}$,  will therefore be the one which minimizes $F(\{V\}, T)$ at a given fixed temperature $T$. By comparison, the static equilibrium lattice, or bare lattice, $\{V^0\}$, minimizes the lattice-dependent static total energy, $E_{\textrm{stat}}^{\textrm{e}}(\{V\})$. The 
zero-point lattice modification can then be defined as the difference between the $T=0$~K and bare lattices, \mbox{$\{V(T=0)\} - \{V^0\}$}.

At this point, we emphasize that, for a general crystal, the temperature-dependent lattice structure, $\{V(T)\}$, includes, in principle, the internal coordinates. While this point is briefly addressed in Ref.~\cite{allen_quasi-harmonic_2019}, it is most customary in the literature to neglect this degree of freedom when evaluating Eq.~(\ref{eq:ftot-helm-full}). It is actually absorbed inside the sampled lattice structures: the internal coordinates are fully optimized in the static $T=0$~K DFT framework, for the imposed $T$-dependent lattice parameters and angles. The phonon frequencies entering Eq.~(\ref{eq:fvib-vol}) are also evaluated at these static optimized configurations. In that case, the FE becomes
\begin{equation}\label{eq:feapprox}
\begin{split}
    F(\{V\}, T)&\simeq F(\{a_i\}, \{b_i\}, T)\\
    &\triangleq F(\{a_i\}, \{b_i\}, \{z_i|\textrm{min}\,E^{\textrm{e}}_{\textrm{stat}}(\{a_i, b_i\})\}, T).
    \end{split}
\end{equation}

We then define the temperature-dependent lattice structure $\{\tilde{V}(T)\}$ as the set of lattice parameters and angles, $\{\tilde{a}_i(T)\}$ and $\{\tilde{b}_i(T)\}$,  which minimizes $F(\{a_i\}, \{b_i\}, T)$ at fixed temperature $T$, as well as the internal coordinates $\{\tilde{z}_i(T)\}$ which minimize the static DFT electronic energy at this lattice structure:
\begin{equation}\label{eq:latticeapprox}
\begin{split}
    \{\tilde{V}(T)\} = \{\; &\{\tilde{a}_i(T), \tilde{b}_i(T)\}|\textrm{min}\,F(\{a_i\}, \{b_i\}, T),\\ &\{\tilde{z}_i(T)\}|\textrm{min}\,E^{\textrm{e}}_{\textrm{stat}}(\{\tilde{a}_i(T), \tilde{b}_i(T)\})\;\}.
\end{split}
\end{equation}
For materials with axial symmetry, $F(\{V\}, T)$ thus reduces to a 2D surface, $F(a, c, T)$, such that the ZPLE is given by \mbox{$\{a(T=0)-a^0,\,c(T=0)-c^0\}$}. For cubic symmetry, one recovers a 1D FE curve, $F(V,T)$, with linear \mbox{ZPLE $=a(T=0)-a^0$}.

\subsubsection{\gru formalism}\label{sec:theo-gru}
Another approach, known as the \gru formalism, is frequently used in the literature to obtain the temperature-dependent lattice parameters. For simplicity, we first derive the expressions for the ZPLE and thermal expansion coefficients for cubic lattices. In that case, Eq.~(\ref{eq:ftot-helm-full}) reduces to a single dependency on the volume, $\{V\}\rightarrow V$.

We start from the equation of state relating pressure to the Helmholtz FE:
\begin{equation}\label{eq:p-eos}
    P(V,T) = -\left.\frac{\partial F}{\partial V}\right|_{T}.
\end{equation}
 
$F$ can be Taylor-expanded to lowest order around the static equilibrium volume, $V^0$, yielding~\cite{grimvall_thermophysical_1986} 
\begin{equation}\label{eq:pvib-vol}
\begin{split}
    P(V, T)\approx &-\frac{B_0}{V^0}\left(V-V^0\right)\\ &+  \frac{1}{V^0}\sum\limits\qv\gamma^V\qv\wqv(V^0)\left(n\qv(V^0,T)+\frac{1}{2}\right),
    \end{split}
\end{equation}
where $B_0$ is the isothermal bulk modulus, $n\qv$, the Bose-Einstein occupation function of the phonon population, and where we have introduced the volumic mode \gru parameters:
\begin{equation}
    \gamma\qv^V = - \frac{V^0}{\wqv(V^0)}\left.\frac{\partial\wqv}{\partial V}\right|_{V^0}.
\end{equation}
The average of these mode parameters weighted by the phonon contribution to the specific heat, $c\qv(T)$, defines the temperature-dependent bulk \gru constant,
\begin{equation}\label{eq:bulkgrun}
    \gamma (T) = \frac{\sum\qv\gamma^V\qv c\qv(T)}{\sum\qv c\qv(T)},
\end{equation}
often used in the experimental literature to parametrize the thermal expansion coefficient of cubic materials~\cite{grimvall_thermophysical_1986,kagaya_mode_1986}.

The equilibrium configuration at any given temperature will verify $P=0$. We can then express the relative volume change with respect to static equilibrium at temperature $T$ as
\begin{equation}\label{eq:deltav}
    \frac{\Delta V (T)}{V^0} = \frac{1}{B_0V^0}\sum\limits\qv\gamma^V\qv\wqv\left(n\qv(T) + \frac{1}{2}\right),
\end{equation}
where from now on we simplify the notation,  \mbox{$\wqv(V^0)\rightarrow\wqv$} and \mbox{$n\qv(V^0,T)\rightarrow n\qv(T)$}.

Evaluating Eq.~(\ref{eq:deltav}) at $T=0$~K yields the volumic ZPLE:
\begin{equation}\label{eq:isotropic-zple}
    \Delta V (T=0) =\frac{1}{B_0}\sum\limits\qv\gamma^V\qv\frac{\wqv}{2}.
\end{equation}
The \gru formalism as presented in most textbooks neglects the zero-point energy in the evaluation of Eq.~(\ref{eq:pvib-vol}), as it focuses on the thermal expansion coefficient, hence the absence of zero-point correction to the lattice volume.

We now proceed to generalize the \gru formalism to anisotropic materials, yet still retaining the previously defined approximation for the internal atomic coordinates (see Eq.~(\ref{eq:feapprox}) and the second line of Eq.~(\ref{eq:latticeapprox}), with $\{\tilde{a}_i(T)\}$ now referring to the temperature-dependent lattice parameters obtained from the \gru approach). To simplify the notation, we will refer to $\{\tilde{a}_i(T)\}$ as $\{a_i(T)\}$.
Until now, we have defined the lattice structure in terms of the $\{a_i\}$ and  $\{b_i\}$, which are well defined regardless of any reference structure. This is a natural choice when working with the FE. As the \gru approach is based on an expansion around the static equilibrium configuration, we will rather use the strain tensor, $\bm{\epsilon}$, whose components (here expressed in Voigt notation~\cite{grimvall_thermophysical_1986}) describe the relative deformations of the lattice parameters and lattice angles with respect to a given reference configuration, which we choose to be the static equilibrium structure:
 
\begin{equation}\label{eq:strains}
\begin{split}
    \epsilon_{i=1,2,3} &= \frac{a_i - a_i^0}{a_i^0},\\
    \epsilon_{i=4,5,6} &= \frac{b_i - b_i^0}{b_i^0}.\\
    \end{split}
\end{equation}
Note that other definitions of the reference configuration are commonly used in the literature, such as the room temperature and the dynamical $T=0$~K (including zero-point effects) structures. These are natural choices when reporting experimental values of the thermal expansion coefficients.

We start from the differential form of the Helmholtz FE~\cite{grimvall_thermophysical_1986},
\begin{equation}\label{eq:helmholtzstrain}
    dF = -k_B\text{ln}Z^{\textrm{ph}}dT + V^0\sum\limits_{i=1}^6 \sigma_i d\epsilon_i.
\end{equation}
The stress tensor, $\bm{\sigma}$, is related to the strain tensor, $\bm{\epsilon}$, by the elastic stiffness tensor, $\bm{c}$, whose components $c_{ij}$ are the isothermal elastic constants of the material,
\begin{equation}
    \sigma_i = \sum\limits_{j=1}^6 c_{ij}\epsilon_j.
\end{equation}

From Eq.~(\ref{eq:helmholtzstrain}), we obtain the generalization of Eq.~(\ref{eq:p-eos}) to anisotropic materials by replacing the volume derivative by derivatives with respect to a given strain. Note that the sign of the derivative is now positive as the components $\sigma_i$ describe the internal stresses:
\begin{equation}\label{eq:sigma-eos}
    \sigma_i(T) = \frac{1}{V^0}\frac{\partial F(T)}{\partial \epsilon_i}.
\end{equation}
The generalization of Eq.~(\ref{eq:pvib-vol}) is again obtained by Taylor-expanding around the equilibrium configuration at zero strain:
\begin{equation}\label{eq:pvib-gen}
    \begin{split}
        \sigma_i&\approx \frac{1}{V^0}\left[V^0 \sum\limits_{j=1}^6 c_{ij}\epsilon_j\right.\\
        &-\left.\sum\limits\qv \gamma^i\qv\wqv(\bm{\epsilon}=0)\left(n\qv(\bm{\epsilon}=0,T) + \frac{1}{2}\right)\right],
    \end{split}
\end{equation}
where we introduce the mode \gru tensor
\begin{equation}\label{eq:mode-gru-tensor}
    \gamma^i\qv = -\frac{1}{\wqv}\left.\frac{\partial\wqv(\epsilon_i)}{\partial\epsilon_i}\right|_{\bm{\epsilon}=0},
\end{equation}
in which the derivative are now taken with respect to the individual strains, $\epsilon_i$.

From now on, we simplify the notation by assuming that the phonon frequencies and occupation function are evaluated at zero strain. Further assuming that the stress components are all zero at the equilibrium configuration of a given temperature, this reduces to 
\begin{equation}\label{eq:sumstrain}
    \sum_{j=1}^6c_{ij}\epsilon_j (T) = \frac{1}{V^0} \sum\qv \gamma^i\qv\wqv\left(n\qv(T) + \frac{1}{2}\right).
\end{equation}
In order to obtain the generalized form of Eq.~(\ref{eq:deltav}), we take the tensor product with the elastic compliance tensor, $\bm{s}$:
\begin{equation}
        \sum\limits_{i=1}^6s_{ki}\sum\limits_{j=1}^6 c_{ij}\epsilon_j = \frac{1}{V^0}\sum\limits_{i=1}^6s_{ki} \sum\qv \gamma^i\qv\wqv\left(n\qv(T) + \frac{1}{2}\right).
\end{equation}
The left-hand side of this expression reduces to the individual strains as the elastic compliance tensor is, by definition, the inverse of the elastic stiffness tensor. At finite temperatures, one could further refine Eq.~(\ref{eq:sumstrain}) and use temperature-dependent elastic constants, which can be obtained by differentiating the Helmholtz FE at temperature $T$ with respect to strains~\cite{malica_quasi-harmonic_2020}. 

For any anisotropic material, the temperature-dependent strains within the QHA can therefore be expressed as 
\begin{equation}\label{eq:deltav-strain}
    \epsilon_k(T) = \frac{1}{V^0}\sum\qv \sum_{i=1}^6 s_{ki}\gamma\qv^i\wqv \left(n\qv(T) + \frac{1}{2}\right).
\end{equation}

For linear thermal expansion of the material along the crystallographic axis, the relevant strains are ${\epsilon_k(T) = \Delta a_k(T)/a_k^0}$, where ${k=1,2,3}$ are Voigt indices. 
Recalling that the thermal expansion coefficients, $\alpha_k$, are the temperature derivatives of the strains at constant stress,
\begin{equation}\label{eq:alpha-derivative}
    \alpha_k = \left.\frac{d\epsilon_k}{dT}\right|_{\sigma},
\end{equation}
we obtain~\cite{barron_thermal_1982}
\begin{equation}\label{eq:alpha-strain}
    \begin{split}
    \alpha_k (T) &= \frac{1}{V^0}\sum\qv \sum_{i=1}^6 s_{ki}\gamma\qv^i \frac{d}{dT}(\wqv n\qv(T)+\frac{1}{2})\\
     &= \frac{1}{V^0}\sum\qv \sum_{i=1}^6 s_{ki}\gamma\qv^i \frac{d}{dT}(\tilde{\varepsilon}\qv(T))\\
    &= \frac{1}{V^0}\sum\qv \sum_{i=1}^6 s_{ki}\gamma\qv^i c^V\qv(T),
    \end{split}
\end{equation}
where $\tilde{\varepsilon}\qv(T)$ is the average phonon energy at temperature $T$ and $c^V\qv(T)$ is the phonon specific heat at constant volume at this temperature.

Equation~(\ref{eq:alpha-strain}) thus generalizes the \gru approach to any material, under our approximation for the internal coordinates. It has the exact same form as if one would have neglected the zero-point energy in Eq.~(\ref{eq:deltav-strain}), as is typically done within this formalism. Nevertheless, if we evaluate Eq.~(\ref{eq:deltav-strain}) at $T=0$~K, we obtain a generalized expression for the ZPLE: 
\begin{equation}\label{eq:zpr-latt-grun}
    \begin{split}
     \Delta a_k (T=0)
    &= \frac{a_k^0}{V^0}\sum\qv \sum_{i=1}^6 s_{ki}\gamma\qv^i\frac{\wqv}{2} .
    \end{split}
\end{equation}
Note that, for the remainder of this article, the ZPLE refers to the zero-point length shift of the lattice parameters only.  In principle, one could also extract from Eq.~(\ref{eq:deltav-strain}) the zero-point variation of the cell angles, for materials in which they are not set to zero by symmetry.

By evaluating the temperature-dependent lattice constants as
\begin{equation}\label{eq:aoft-strain}
    a_k (T) = a_k^0\int\limits_0^T\; \alpha_k(T')\;dT' + a_k\,(T=0),
\end{equation}
where $a_k(T=0)=a_k^0 + \Delta a_k (T=0)$ is the ZPLE-renormalized lattice constant obtained from Eq.~(\ref{eq:zpr-latt-grun}), the zero-point lattice effect can thus easily be taken into account within the anisotropic \gru formalism. To the best of our knowledge, no ZPLE results were explicitly reported in the literature using this formula. We note that this expression can be seen as the anisotropic generalization of Eq.~(4) of Ref.~\cite{huang_efficient_2016}, in the $T=0$~K, $P=0$, one-shot limit. The isotropic Eq.~(\ref{eq:isotropic-zple}) was also used by Querales-Flores \textit{et al.}~\cite{querales-flores_temperature_2019} to report the ZPLE contribution to the ZPR$_g$ in cubic PbTe. See also Sec. \MakeUppercase{\romannumeral 4} of Ref.~\cite{allen_quasi-harmonic_2019} and references therein.

Section~S1
of the Supplemental Material~\cite{note_supplemental_material} extends the \gru formalism to a fixed external hydrostatic pressure, while Sec.~S2 shows how the well-known literature results for the case of cubic and axial symmetries can be easily retrieved from Eq.(\ref{eq:zpr-latt-grun}). We state their explicit ZPLE expressions and show that the isotropic Eq.~(\ref{eq:isotropic-zple}) is contained in the anisotropic formulation of Eq.~(\ref{eq:deltav-strain}).

\subsection{Estimation of ZPLE correction to the band gap energy}

We now estimate the size of the ZPLE correction to the electronic eigenenergies, $\varepsilon\kn$, induced by the zero-point motion in terms of simple parameters. In the following, we will first retain the temperature dependence of the electronic eigenenergies, starting from the volumic simplification of the TE contribution in Eq.~(\ref{eq:thermo-general}) and replacing the partial derivatives with respect to volume by logarithmic derivatives:
\begin{equation}\label{eq:thermo-cubic}
    \left(\frac{\partial\varepsilon\kn(V,T)}{\partial T}\right)_P^{\textrm{TE}} =  \left(\frac{\partial\varepsilon\kn(V,T)}{\partial \text{ln}V}\right)_T\left(\frac{\partial\text{ln}V}{\partial T}\right)_P,
\end{equation}
then take the $T\rightarrow0$~K limit to obtain the final expression. We stress that, in principle, Eq.~(\ref{eq:thermo-cubic}) should only be applied to cubic systems. Nevertheless, since we are seeking for an approximate expression, it can be expected to be reasonably accurate for sufficiently symmetric anisotropic materials, such as the wurtzite structure.

From the chain rule, we can recast the constant temperature term in terms of a derivative with respect to pressure and the bulk modulus, $B_0$:
\begin{equation}\label{eq:dedpb0}
    \begin{split}
        \frac{\partial\varepsilon\kn}{\partial\text{ln}V} &= \frac{\partial\varepsilon\kn}{\partial P}\frac{\partial P}{\partial\text{ln}V}\\
        &= \frac{\partial\varepsilon\kn}{\partial P}(-B_0).
    \end{split}
\end{equation}
By definition, the constant pressure term is simply the volumic thermal expansion coefficient, $\beta(T)$:
\begin{equation}\label{eq:beta}
    \begin{split}
        \frac{1}{V}\frac{\partial V}{\partial T} &= \beta(T)\\
        &= \frac{\partial}{\partial T}\left(\frac{\Delta V(T)}{V^0}\right).
    \end{split}
\end{equation}
This expression can be adapted to any crystal symmetry by recalling that $\beta(T)$ is the trace of the thermal expansion tensor, whose coefficients describe the strain induced on the lattice by an infinitesimal temperature increment:
\begin{equation}\label{eq:beta-derivative}
    \begin{split}
        \beta(T) &= \sum\limits_{k=1}^3 \alpha_k (T),\\
        &= \sum\limits_{k=1}^3 \left(\frac{\partial\epsilon_k}{\partial T}\right).
        \end{split}
\end{equation}
In the case of cubic and axial symmetry, this reduces to
\begin{equation}\label{eq:beta-derivative-cases}
        \beta(T) = \begin{cases}
        3\,\alpha(T),\qquad &\text{(cubic),}\\
         2\,\alpha_a(T) + \alpha_c(T),\qquad &\text{(axial)}.
         \end{cases}
\end{equation}
Substituting Eq.~(\ref{eq:dedpb0}) and (\ref{eq:beta}) in the last term of Eq.~(\ref{eq:thermo-cubic}), we can express the rate of change of the electronic eigenvalues with respect to temperature at any temperature $T'$ as 
\begin{equation}
        \left(\frac{\partial\varepsilon\kn(V,T)}{\partial T}\right)_P^{\textrm{TE}} (T')= 
     -\frac{\partial\varepsilon\kn}{\partial P}\, B_0(T')\beta(T').
\end{equation}
The thermal expansion-induced change to the electronic eigenenergies is obtained by integrating the previous expression from $T'=0$ to $T'=T$, recalling that the lower bound is the dynamical $T=0$~K lattice. Further neglecting the temperature dependence of the bulk modulus, which is typically very small at low temperatures, and taking advantage of the definition of $\beta(T')$ as a temperature derivative, we obtain
\begin{equation}
\begin{split}
    \Delta\varepsilon\kn (T)&= \varepsilon\kn(T) - \varepsilon\kn^0\\
    &\approx -B_0\frac{\partial\varepsilon\kn}{\partial P}\left(\frac{\Delta V(T)}{V^0} + \frac{\Delta V}{V^0}^{\textrm{ZPLE}}\right),
    \end{split}
\end{equation}
 where the integration constant (second term in the large parentheses) captures the $T=0$~K volume variation compared to the static lattice and is equivalent to the second integration constant in Eq.~(\ref{eq:eigcorr}).
Hence, 
one obtains the ZPLE-induced modification of an electronic eigenstate,
\begin{equation}\label{eq:ZPLE-band_estimation}
    \Delta\varepsilon\kn (T=0)\approx -B_0\frac{\partial\varepsilon\kn}{\partial P}\frac{\Delta V}{V^0}^{\textrm{ZPLE}},
\end{equation}
where $\Delta V^{\textrm{ZPLE}}/V^0= (V(T=0)-V^0)/V^0$. In the following, we will refer to this term as $\Delta V/V^0$ for simplicity.

We finally note that, since the volume of the unit cell is allowed to change, the interatomic distances are 
modified, which in turn affects the ionic potential, hence the value of the Fermi energy. To remove any ambiguity regarding the reference energy of a given electronic band, we use Eq.~(\ref{eq:ZPLE-band_estimation}) to compute directly the change to the band gap energy, which is defined irrespective of a reference Fermi energy.

The first two factors entering the right-hand side of Eq.~(\ref{eq:ZPLE-band_estimation}), namely, the bulk modulus and the band gap derivative with respect to pressure (also known in the literature as the pressure coefficient), can easily be obtained either from experimental data or from static DFT calculations. The last one, $\Delta V/V^0$, requires a little more care and can be estimated in two ways. On the one hand, one can extrapolate the high-temperature limit slope of the curve for $\Delta a(T)/a^0$ and calculate its intercept at $T=0$~K. The ZPLE is then simply the difference between the extrapolated value and the actual lattice parameter at $T=0$ (see Fig.~2 of Ref.~\cite{cardona_isotope_2005}, and note that our definition of $\Delta a/a^0$ differs from theirs by a sign change). This procedure is similar to the one frequently used in the literature to evaluate the full band gap ZPR from experimental values of $E_{g}(T)$ (see Fig.~21 of Ref.~\cite{cardona_isotope_2005}). Note that the accuracy of such extrapolation depends on the measurements extending to sufficiently high temperature, i.e. much larger than the Debye temperature, as will be discussed in Sec.~\ref{sec:res-estimate}. One could also use theoretical results for $\Delta V/V^0$ from Eq.~(\ref{eq:deltav-strain}) to validate Eq.~(\ref{eq:ZPLE-band_estimation}) a posteriori.

Lastly, one could also, in principle, make use of a simple semiempirical model proposed in the literature to express $\Delta V/V^0$ in terms of a few parameters~\cite{alchagirov_energy_2001}. This model relies on the Debye model for the expression for the zero-point energy and the Dugdale-MacDonald parametrization of the bulk \gru constant in terms of the first derivative of the bulk modulus with respect to pressure, $B_1$~\cite{dugdale_thermal_1953}. Other parametrizations of $\gamma^V$ in terms of $B_1$ have been proposed (see Ref.~\cite{lejaeghere_ab_2014} and references therein). 

This simple model was widely used in the early 2000s to account for the ZPLE when comparing the accuracy of lattice parameter prediction from different exchange-correlation functionals (see Ref.~\cite{hao_lattice_2012} and references therein) and was originally derived for crystals with one atom in the unit cell. Ref.~\cite{haas_calculation_2009} showed that it strongly overestimates the ZPLE for diamond and zincblende structures. In that case, the poor predictive capabilities of the semiempirical model can be understood by recalling that it is based on bulk properties. Namely, it supposes that the average behavior of the whole phonon population can be approximated by the behavior of a single \enquote{average phonon}, whose frequency is lowered by increasing pressure. For most diamond and zincblende materials, the thermal expansion coefficient is negative at low temperatures. 
This behavior is driven by low-frequency modes, which are typically associated with bending modes that strengthen the atomic bonds during volumic expansion~\cite{xu_theory_1991}, hence a phonon hardening.

By looking at Eq.~(\ref{eq:bulkgrun}), we can understand that the proper definition of the bulk \gru constant as a weighted sum of the individual \gru mode parameters will correctly capture this unusual behavior, as it allows each mode parameter to be either positive or negative. Ignoring these alternating signs for low-frequency modes, as is done in the semiempirical model, naturally yields an overestimation of the total bulk \gru constant. 
We will thus restrict our estimation of $\Delta a_k(T=0)/a_k^0$ to the high-temperature slope method explained earlier.
\FloatBarrier
\subsection{Electron-phonon interaction}\label{sec:theory-epi}
\begin{figure}[t]
    \centering
    \includegraphics[width=\columnwidth]{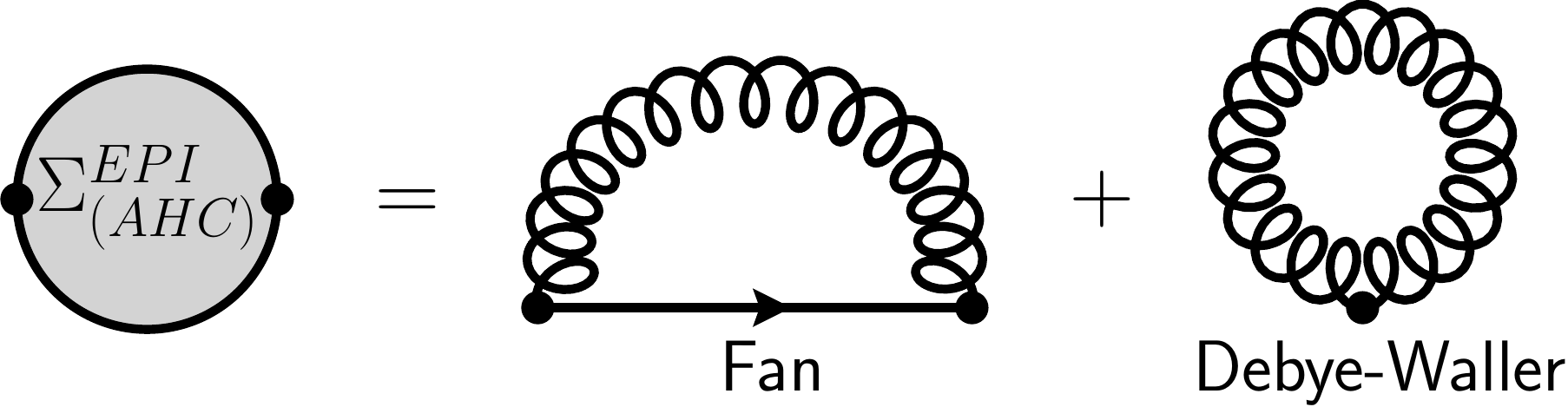}
    \caption{Fan and Debye-Waller contributions entering the lowest order EPI self-energy in the AHC framework.}
    \label{fig:fandw}
\end{figure}
The EPI contribution to the temperature dependence of the band gap energy, and, consequently, on $\textrm{ZPR}_g$, has been thoroughly discussed in the literature. Many theoretical formalisms have been used to address this question, among them ab initio molecular dynamics~\cite{franceschetti_first_2007, ramirez_path_2008}, path integral molecular dynamics~\cite{kundu_quantum_2021}, importance sampling Monte Carlo~\cite{monserrat_temperature_2014}, finite differences and distorted supercells~\cite{antonius_dynamical_2015, monserrat_comparing_2014, monserrat_electronphonon_2018}, the special displacement method~\cite{zacharias_theory_2020}, the Bethe-Salpeter equation for excitonic gaps~\cite{marini_abinitio_2008} as well as the Allen-Heine-Cardona (AHC) formalism~\cite{allen_temperature_1983, allen_theory_1976, allen_theory_1981}, either adiabatic~\cite{giustino_electron-phonon_2010, ponce_temperature_2014} or non-adiabatic~\cite{miglio_predominance_2020, lihm_phonon-phonon_2020}. In the following, we briefly review the key concepts of the non-adiabatic AHC framework. We invite our readers to consult Refs.~\cite{giustino_electron-phonon_2017, ponce_temperature_2014, ponce_temperature_2015} for a complete discussion of this methodology and of its underlying approximations.

In many-body perturbation theory~\cite{mahan_many-particle_1990}, the zero-point renormalization of an electronic energy level $\varepsilon\kn$ is captured by the real part of the electron-phonon self-energy, $\Sigma\kn^{\text{EPI}}$, evaluated at the quasiparticle energy, $\varepsilon\kn(T=0)$:
\begin{equation}
\begin{split}
    \textrm{ZPR}\kn^{\textrm{EPI}} &= 
    \varepsilon\kn(T=0) - \varepsilon^0\kn \\  &= \mathfrak{Re}\left[\Sigma\kn^{\textrm{EPI}}(\omega=\varepsilon\kn(T=0), T=0)\right].
    \end{split}
\end{equation}
In this expression, $\varepsilon^0\kn$ is the unperturbed Kohn-Sham (KS) eigenvalue computed at the fixed lattice geometry which minimizes the total electronic energy of the system in the static approximation, $\{V^0\}$. For the remainder of this section, $T=0$ is implied throughout.

The self-energy at second order in perturbation theory can be expressed as the sum of the Fan and Debye-Waller contributions, shown in Fig.~\ref{fig:fandw}. Assuming that the quasiparticle energy is close to the bare eigenenergy, one can evaluate the frequency-dependent Fan self-energy at the unperturbed electronic energy, in the so-called on-the-mass-shell approximation~\cite{marini_many-body_2015}, yielding
\begin{equation}
\textrm{ZPR}\kn^{\textrm{EPI}}  \approx \mathfrak{Re}\left[\Sigma\kn^{\textrm{Fan}}(\omega=\varepsilon^0\kn) + \Sigma\kn^{\textrm{DW}}\right].
\end{equation}
We finally approximate the fully-interacting wavefunction by the KS-DFT wavefunction and obtain the standard expression for the non-adiabatic Fan self-energy~\cite{ponce_temperature_2014, giustino_electron-phonon_2017}:
\begin{equation}\label{eq:sigmafan}
    \begin{split}
   & \Sigma_{\mathbf{k}n}^{\text{Fan}}(\varepsilon\kn^0) = 
   \sum\limits\qv^{\text{BZ}}\sum\limits_{n'}
   |\bra{\mathbf{k+q}n'}V^{(1)}_{\mathbf{q}
   \nu}\ket{\mathbf{k}n}|^2 \times \\
   & \left[ \frac{1 -
   f_{\mathbf{k+q}n'}}{\varepsilon\kn^0-\varepsilon^0\subkn{k+q}{n'}-
   \wqv + i\eta_\mathbf{k}} + \frac{
   f_{\mathbf{k+q}n'}}{\varepsilon^0\kn-\varepsilon^0_{\mathbf{k+q}n'}+
   \wqv + i\eta_\mathbf{k}} \right].
    \end{split}
\end{equation}
The electron-phonon matrix elements, \mbox{$\bra{\mathbf{k+q}n'}V^{(1)}_{\mathbf{q}
   \nu}\ket{\mathbf{k}n}$}, capture the probability amplitude of a coupling between two electronic eigenstates, given the first-order change in the KS potential induced by the collective atomic motion along the $\mathbf{q}\nu$-phonon mode, $V^{(1)}_{\mathbf{q}\nu}$. Since we work at $T=0$~K, the two terms inside the square brackets can be interpreted as virtual phonon emission processes by respectively electrons and holes~\cite{coleman_introduction_2015}. In the case of semiconductors, the Fermi-Dirac occupation functions at $T=0$~K, $f_{\mathbf{k+q}n'}$, are either 0 for conduction bands or 1 for occupied states. The small parameter $\eta_\mathbf{k}$ in the denominators correctly shifts the poles of the Green's function in the complex plane to maintain causality.

The first-order potential can be written as 
\begin{equation}\label{eq:v1-with-operator}
\begin{split}
V^{(1)}_{\mathbf{q}
   \nu} &= \frac{1}{\sqrt{2\wqv}}\sum\limits_{\kappa\alpha}U_{\nu, \kappa\alpha}(\mathbf{q}) \sum\limits_l \e^{i\mathbf{q}\cdot\bm{R}_l}\frac{\partial V^{\rm{KS}}}{\partial \bm{R}_{l\kappa\alpha}}  \\ 
   &\triangleq \nabla\qv V^{\rm{KS}},
 \end{split}
\end{equation}
where $\bm{R}_{l\kappa\alpha}$ represents the displacement of atom $\kappa$, located in unit cell $l$, in cartesian direction $\alpha$, and $V^{\rm{KS}}$ is the Kohn-Sham potential. The second line emphasizes the definition of the $\nabla\qv$ operator. The phonon eigendisplacement vectors, $U_{\nu, \kappa\alpha}(\mathbf{q})$, are the solutions of the generalized eigenvalue equation 
\begin{multline}
M_\kappa \omega\qv^2U_{\nu,\kappa\alpha}(\mathbf{q}) = \\ \sum\limits_{\kappa' \alpha'}\left(\sum\limits_l \e^{i\mathbf{q}\cdot\bm{R}_l}\frac{\partial^2 E}{\partial\bm{R}_{l\kappa\alpha}\partial\bm{R}_{0\kappa'\alpha'}}\right) U_{\nu, \kappa' \alpha'}(\mathbf{q}), 
\end{multline}
where $M_\kappa$ is the atomic mass of atom $\kappa$ and $E$ is the total energy of the static lattice. They verify the normalization condition
\begin{equation}
    \sum\limits_{\kappa\alpha}M_\kappa U_{\nu,\kappa\alpha}^\ast(\mathbf{q})U_{\nu',\kappa\alpha}(\mathbf{q}) = \delta_{\nu\nu'}.
\end{equation}

The static Debye-Waller self-energy formally involves second derivatives of the KS potential evaluated at first order in perturbation theory:
\begin{equation}
    \Sigma^{\text{DW}}\kn =
\sum\limits\qv\frac{1}{2}\bra{\mathbf{k}n}
\nabla\qv \nabla_{-\mathbf{q}\nu}V^{\rm{KS}} \ket{\mathbf{k}n}.
\end{equation}
However, in the rigid-ion approximation (RIA), the translational invariance of the lattice allows the second-order matrix elements to be recast in terms of the first derivatives entering the Fan self-energy~\cite{ponce_temperature_2014}. By doing so, we suppose that the second derivatives of the Kohn-Sham potential with respect to the displacement of two different atoms within the unit cell vanish. The validity of this approximation was verified for crystals~\cite{ponce_temperature_2014, antonius_manybody_2014}, but strongly questioned for diatomic molecules~\cite{gonze_theoretical_2011}.

\section{\label{sec:computation}Computational details}
\subsection{Ground state calculations and lattice optimization}\label{sec:gsopt}
All first-principles calculations were performed with the \textsc{Abinit} software package~\cite{gonze_abinitproject:_2019}. We obtain the static ground state electronic structure and total energy at $T=0~$K from density functional theory (DFT). Spin-orbit coupling is neglected throughout. For most materials, we use norm-conserving pseudopotentials from the PseudoDojo project~\cite{van_setten_pseudodojo:_2018} and rely on the generalized gradient approximation of the Perdew-Burke-Ernzerhof functional (PBE-GGA)~\cite{perdew_generalized_1996}. Lattice parameters and relevant internal atomic coordinates were optimized until all forces on the atoms were below \mbox{$10^{-5}$ hartree/Bohr$^3$}.
The optimized lattice parameters were used throughout, except for Ge, as the optimized structure led to an unphysical gapless band structure, and GaP, to ensure the correct relative energy ordering of the electron valleys in the conduction band. In those cases, we used the experimental lattice parameter extrapolated to $T=0$~K. This was done to ensure a correct evaluation of the EPI contribution to $\textrm{ZPR}_g$, which can be very sensitive to such details.\\

For BN and AlN, the reported values for the EPI contribution to $\textrm{ZPR}_g$ comes from previously published data~\cite{ponce_temperature_2015}. Hence, for consistency, we used for these materials the Perdew and Wang parametrization of the local density approximation (LDA)~\cite{perdew_accurate_1992}, FHI98 pseudopotentials~\cite{FUCHS199967} and the reported lattice parameters for those materials. The energy cutoffs, optimized lattice parameters and \kpoint sampling of the Brillouin zone (BZ) used for all materials can be found in Table~S1
of the Supplemental Material~\cite{note_supplemental_material}.

\subsection{Lattice dynamics and electron-phonon interaction}\label{sec:computation-epi}
Response-function and electron-phonon coupling calculations were performed within the density functional perturbation theory (DFPT) framework~\cite{gonze_dynamical_1997}. The \qpoint sampling used to compute the phonon frequencies and electron-phonon matrix elements for the different materials can be found in Table~S1
of the Supplemental Material~\cite{note_supplemental_material}. We used an imaginary broadening of $\eta=0.01$~eV in the denominator of Eq.~(\ref{eq:sigmafan}) and accelerated the convergence with respect to the \qpoint sampling using the methodology based on the Fr\"ohlich model presented in the Supplementary Methods of Ref.~\cite{miglio_predominance_2020}. The Sternheimer equation method is used to reduce the explicit number of bands required in the active subspace of the self-energy~\cite{gonze_theoretical_2011}. The explicit number of bands was chosen such that the energy difference between the CBM and the highest band was at least 20 eV. Most of the $\text{ZPR}_g^{\text{EPI}}$ results used in this paper were previously reported in Ref.~\cite{miglio_predominance_2020}. 
For GaAs, the electron-phonon matrix elements for the maximal $\mathbf{q}$-point grid  listed in Table~S1 of the Supplemental Material~\cite{note_supplemental_material} were obtained by interpolating the first-order potential computed self-consistently on a \mbox{$16\times16\times16$} \mbox{$\mathbf{q}$-point} grid, using the methodology described in Ref.~\cite{gonze_abinitproject:_2019}.
\subsection{Zero-point lattice expansion and band gap correction}
For both the FE minimization and \gru methodologies, we evaluated the vibrational spectrum of all lattice configurations using an {8$\times$8$\times$8} {$\Gamma$-centered} \qpoint grid. We verified the convergence of our sampling by comparing with a {24$\times$24$\times$24} Fourier-interpolated \qpoint grid for both approaches. Note that the \qpoint grid used for ZPLE is coarser than the one used for EPI, since it only requires the phonon frequencies, which converge much faster with respect to the \qpoint sampling.\\

For the \gru approach, the relaxed-ion compliance tensor was computed from the strain-strain derivatives of the total energy within DFPT~\cite{hamann_metric_2005}, taking into account the relaxation of the atomic coordinates within the unit cell. Note that properly converged elastic constants typically require a small smearing of the plane wave cutoff energy, hence, a slightly higher effective value compared to GS or phonon calculations,  since shear strains modify the shape of the unit cell. The specific values used for the different materials are explicited in Table~S1 of the Supplemental Material~\cite{note_supplemental_material}. The derivatives entering Eq.~(\ref{eq:mode-gru-tensor}), which is used to evaluate Eq.~(\ref{eq:zpr-latt-grun}), were computed from central finite differences using two volumes per degree of freedom and the static equilibrium configuration. To test the stability of the numerical differentiation, we used strained lattice parameters corresponding to a volume change of $\pm0.5\%$ and $\pm1\%$ for the unit cell for each independent crystallographic axis.\\

For the cubic materials, the FE curve was constructed using 9 configurations within a $\pm5\%$ volume change with respect to the static equilibrium volume. The optimal lattice parameter was obtained by fitting a Murnaghan equation of state to the FE curve. For the wurtzite materials, the free energy surface was computed from 27 volumes for GaN, 42 for AlN and 55 for ZnO. The chosen lattice parameters were such that they individually yielded a volume change within $\pm1\%$ of the equilibrium volume. The 2D surface was minimized by fitting a second-order polynomial. Most results from FE minimization were previously reported in Table~II of the Supplementary Information of Ref.~\cite{miglio_predominance_2020}.\\

We obtain the ZPLE-induced band gap ZPR, ZPR$_g^{\textrm{ZPLE}}$, from static DFT at the renormalized lattice parameters. For materials whose internal coordinates are not fixed by symmetry, i.e. the three wurtzite materials, we use the internal coordinates which minimize the Born-Oppenheimer energy at such parameters.\\

In the case of Ge and GaP, as the static equilibrium configuration was set at the experimental lattice parameter instead of the theoretical one, we minimized the Gibbs free energy with an artificial pressure of respectively 7.88~GPa for Ge and 2.77~GPa for GaP, in order to shift the minimum of the static calculated Born-Oppenheimer energy curve at the experimental lattice parameter.
\section{Results and discussion}\label{sec:results}
\subsection{Zero-point lattice expansion}\label{sec:res-ZPLE}
\begin{table*}[]
    \centering
    \caption{\textbf{Zero-point lattice expansion for cubic materials.} $a^0$ is the theoretical cubic lattice parameter in the static lattice approximation, except for Ge and GaP (see Sec.~\ref{sec:gsopt} for details). The variation of $a^0$ induced by the zero-point motion of the ions, $\Delta a^{\textrm{FE}} (T=0)$, is obtained by minimizing the Helmholtz free energy at $T=0$~K, while $\Delta a^{\textrm{Grun}} (T=0)$ is obtained from the \gru parameters, using Eq.~(\ref{eq:zpr-latt-grun}). The indication ($T=0$) is omitted in the table, for sake of simplicity. The central finite differences entering the mode \gru parameters were evaluated with a volume variation of resp. $\pm$1\% and $\pm$0.5\%. The three rightmost columns show the fractional ZPLE with respect to the bare lattice parameter.}
    \label{tab:a0data}
      \setlength\extrarowheight{2pt}
 \begin{tabularx}{\textwidth}{p{20mm}  Y Y >{\hsize=0.7\hsize}Y >{\hsize=0.7\hsize}Y Y >{\hsize=0.7\hsize}Y >{\hsize=0.7\hsize}Y}
    \hline\hline
       \multirow{2}{*}{Material}  &  \multirow{2}{*}{$a^0$ (Bohr)} & \multirow{2}{*}{$\Delta a^{\textrm{FE}}$ (Bohr)} &  \multicolumn{2}{c}{$\Delta a^{\textrm{Grun}}$ (Bohr)} 
       & \multirow{2}{*}{$\Delta a^{\textrm{FE}}/a^0$ (\%)} &  \multicolumn{2}{c}{$\Delta a^{\textrm{Grun}} /a^0$ (\%)}\\
      \cmidrule(lr){4-5}\cmidrule(lr){7-8}
       & & & 1\% & 0.5\% & & 1\% & 0.5\%\\[2pt]
       \hline
        C-dia & ~6.7534 & 0.0253 & 0.0250 & 0.0248&0.375& 0.370&0.367\\
        Si-dia & 10.3348 & 0.0174 & 0.0172 &0.0172& 0.169& 0.166& 0.166\\
        Ge-dia &10.5220 & 0.0102 & 0.0112 & 0.0112&0.097& 0.106&0.106\\
        SiC-zb & ~8.2772 & 0.0227 & 0.0226&0.0223 &0.274& 0.273 &0.269\\
        \hline
        BN-zb & ~6.7460 & 0.0285 & 0.0271 &0.0270 &0.423&0.402&0.400\\
        BAs & ~9.0880 & 0.0238 & 0.0237 & 0.0236 &0.262&0.261& 0.260\\
        AlP-zb & 10.4057 & 0.0180 & 0.0178 &0.0178 &0.173& 0.171&0.171\\
        AlAs-zb & 10.8248 & 0.0166 & 0.0162& 0.0163&0.153&0.150&0.151\\
        AlSb-zb & 11.7621 & 0.0163 & 0.0162 & 0.0161&0.139&0.138&0.137\\
        GaN-zb &  ~8.5984 & 0.0209 &0.0208 &0.0208 &0.243&0.242& 0.242\\
        GaP-zb & 10.2942 & 0.0152 & 0.0156&0.0156 &0.148&0.152&0.152\\
        GaAs-zb & 10.8627 & 0.0157 & 0.0146 & 0.0145& 0.145& 0.134&0.134\\
        \hline
        ZnS-zb & 10.2859 & 0.0194 &0.0193& 0.0192&0.189&0.188& 0.187\\
        ZnSe-zb & 10.8330 & 0.0160 & 0.0157& 0.0155&0.148&0.145&0.143\\
        ZnTe-zb & 11.6819 & 0.0154 & 0.0153& 0.0153&0.132& 0.131&0.131\\
        CdS-zb & 11.2021 & 0.0177 & 0.0170& 0.0162&0.158&0.152&0.145\\
        CdSe-zb & 11.7114 & 0.0135 & 0.0133 & 0.0135&0.115&0.114&0.115\\
        CdTe-zb & 12.5133 & 0.0129 & 0.0126 & 0.0125&0.103&0.101&0.100\\
        MgO-rs & ~8.0366 & 0.0333 & 0.0317& 0.0318&0.414& 0.394& 0.396\\
          \hline\hline
    \end{tabularx}
\end{table*}
\begin{table*}[]
    \centering
    \caption{\textbf{Zero-point lattice expansion for wurtzite materials.} $a^0$ and $c^0$ are the theoretical lattice parameters in the static lattice approximation. The variation of $a^0$ and $c^0$ induced by the zero-point motion of the ions, respectively $\Delta a^{\textrm{FE}} (T=0)$ and $\Delta c^{\textrm{FE}} (T=0)$, are obtained by minimizing the 2D Helmholtz free energy surface, while $\Delta a^{\textrm{Grun}} (T=0)$ and $\Delta c^{\textrm{Grun}} (T=0)$ are obtained from the \gru parameters, using Eq.~(\ref{eq:zpr-latt-grun}). The indication ($T=0$) is omitted in the table, for sake of simplicity. The central finite differences required to obtain the mode \gru parameters were evaluated using a volume variation of resp. $\pm 1\%$ and $\pm0.5\%$. The three rightmost columns show the fractional ZPLE with respect to the bare lattice parameters. The bottom section reports the resulting variation of the $c/a (T=0)$ ratio.}
    \label{tab:a0c0data}
      \setlength\extrarowheight{2pt}
       \begin{tabularx}{\textwidth}{p{20mm}  Y Y >{\hsize=0.7\hsize}Y >{\hsize=0.7\hsize}Y Y >{\hsize=0.7\hsize}Y >{\hsize=0.7\hsize}Y}
    \hline\hline
    \multirow{2}{*}{Material}  &  \multirow{2}{*}{$a^0$ (Bohr)} & \multirow{2}{*}{$\Delta a^{\textrm{FE}}$ (Bohr)} &  \multicolumn{2}{c}{$\Delta a^{\textrm{Grun}}$ (Bohr)} 
      & \multirow{2}{*}{$\Delta a^{\textrm{FE}}/a^0$ (\%)} &  \multicolumn{2}{c}{$\Delta a^{\textrm{Grun}} /a^0$ (\%)}\\
      \cmidrule(lr){4-5}\cmidrule(lr){7-8}
      & & & 1\% & 0.5\% & & 1\% & 0.5\%\\[2pt]
       \hline
       AlN &  5.783 & 0.0170 & 0.0163& 0.0162
        & 0.294 & 0.282 &0.280 \\
       GaN &  6.0841 & 0.0151 & 0.0151 &0.0151 
        & 0.248 & 0.248 &0.248 \\ 
        ZnO &  6.1910 & 0.0165 & 0.0165 &0.0167 & 0.267 & 0.267 &0.270 \\
        \hline
    \multirow{2}{*}{Material}  & \multirow{2}{*}{ $c^0$ (Bohr)} & \multirow{2}{*}{$\Delta c^{\textrm{FE}}$ (Bohr)} &  \multicolumn{2}{c}{$\Delta c^{\textrm{Grun}}$ (Bohr)} 
       & \multirow{2}{*}{$\Delta c^{\textrm{FE}}/c^0$ (\%)} &  \multicolumn{2}{c}{$\Delta c^{\textrm{Grun}} /c^0$ (\%)}\\
       \cmidrule(lr){4-5}\cmidrule(lr){7-8}
       & & & 1\% & 0.5\% & & 1\% & 0.5\%\\[2pt]
          \hline
        AlN &  9.255 & 0.0241 & 0.0242 &0.0242
        & 0.260 & 0.262 &0.262 \\  
        GaN &  9.9110 & 0.0232 & 0.0228 &0.0229
        & 0.234 & 0.230& 0.231 \\
        ZnO &  9.9870 & 0.0213 & 0.0213 &0.0207 
        & 0.213 & 0.214 &0.207 \\
        \hline
    \multirow{2}{*}{Material}  &  \multirow{2}{*}{$c^0/a^0$} & \multirow{2}{*}{$(c/a)^{\textrm{FE}}~(T=0)$} & \multicolumn{2}{c}{$(c/a)^{\textrm{Grun}}~(T=0)$} & \multirow{2}{*}{ $\Delta (c/a)^{\textrm{FE}}$}& \multicolumn{2}{c}{$\Delta (c/a)^{\textrm{Grun}}$}\\
    \cmidrule(lr){4-5}\cmidrule(lr){7-8}
    & & & 1\% & 0.5\% & & 1\% & 0.5\%\\[2pt]
    \hline
    AlN &1.6004 & 1.5998 & 1.6001&1.6001& -0.0006& -0.0003&-0.0003\\
    GaN &1.6290 & 1.6286 & 1.6286& 1.6286& -0.0004 & -0.0004&-0.0004\\
    ZnO & 1.6131 & 1.6122 & 1.6123& 1.6121& -0.0010 & -0.0009&-0.0010\\
          \hline\hline
    \end{tabularx}
\end{table*}
The numerical ZPLE results for the cubic and wurtzite materials are reported in Tables~\ref{tab:a0data} and~\ref{tab:a0c0data} respectively. Note that in most Tables, the material names are appended with their lattice structure, either diamond (dia), zincblende (zb),  rocksalt (rs) or wurtzite (w). As expected, the corrections to the static lattice parameter are small, typically between 0.1-0.2\% of the static value, except for materials containing light atoms (C, B, N, O), where it can reach up to 0.4\%. This can intuitively be understood in terms of a simple mass and spring model, whose vibrational frequency varies as $\sqrt{k/M}$. Recall that the strength of ZPLE depends on how phonon frequencies vary with respect to volume. Lighter materials tend to have a more localized charge density. A small variation of the lattice parameter thus induces a larger charge reorganization, hence a greater variation of the spring constant $k$, compared to heavier materials. Note also that the typical size of the ZPLE is roughly one order of magnitude smaller than the typical error associated with the LDA and PBE functionals, the former usually underestimating the lattice parameter, the latter overestimating it. As ZPLE systematically increases the lattice parameter for our set of materials, it will bring the lattice constant closer to the experimental value for LDA and worsen the agreement for PBE.

In the case of wurtzite materials, our generalized formalism allows for anisotropic ZPLE along the independent crystallographic directions. For the three materials considered, we find that the relative in-plane ZPLE tends to be slightly larger than the out-of-plane ZPLE, resulting in a small, almost negligible decrease of the $c/a$ ratio. This could a posteriori justify using the volumic approach (Eq.~(\ref{eq:deltav})) for this class of materials, therefore approximating the $c/a$ ratio to remain constant. Such a procedure would greatly improve numerical efficiency by reducing the dimensionality of the problem. This might not, however, be advisable for layered or weakly-bonded van der Waals materials, which can display significant elastic anisotropy~\cite{mckinney_ionic_2018}. In that case, as the independent crystallographic axes respond very differently to the zero-point \enquote{phonon pressure}~\cite{munn_role_1972}, the $c/a$ ratio could be more significantly affected even at very low temperatures.

We finally note that even a negligible modification of the $c/a$ ratio at $T=0$~K could lead to a noticeable effect as temperature is increased, should the in-plane and out-of-plane thermal expansion coefficients display sufficiently different behaviors, as can be seen for example in Fig.~(3) of Ref.~\cite{iwanaga_anisotropic_2000}. Hence, for non-cubic materials, the volumic and generalized methodologies can start to deviate even at moderate temperatures.
\subsection{Free energy minimization vs \gru approach}\label{sec:res-compare}
\begin{figure}
    \centering
    \includegraphics[width=\linewidth]{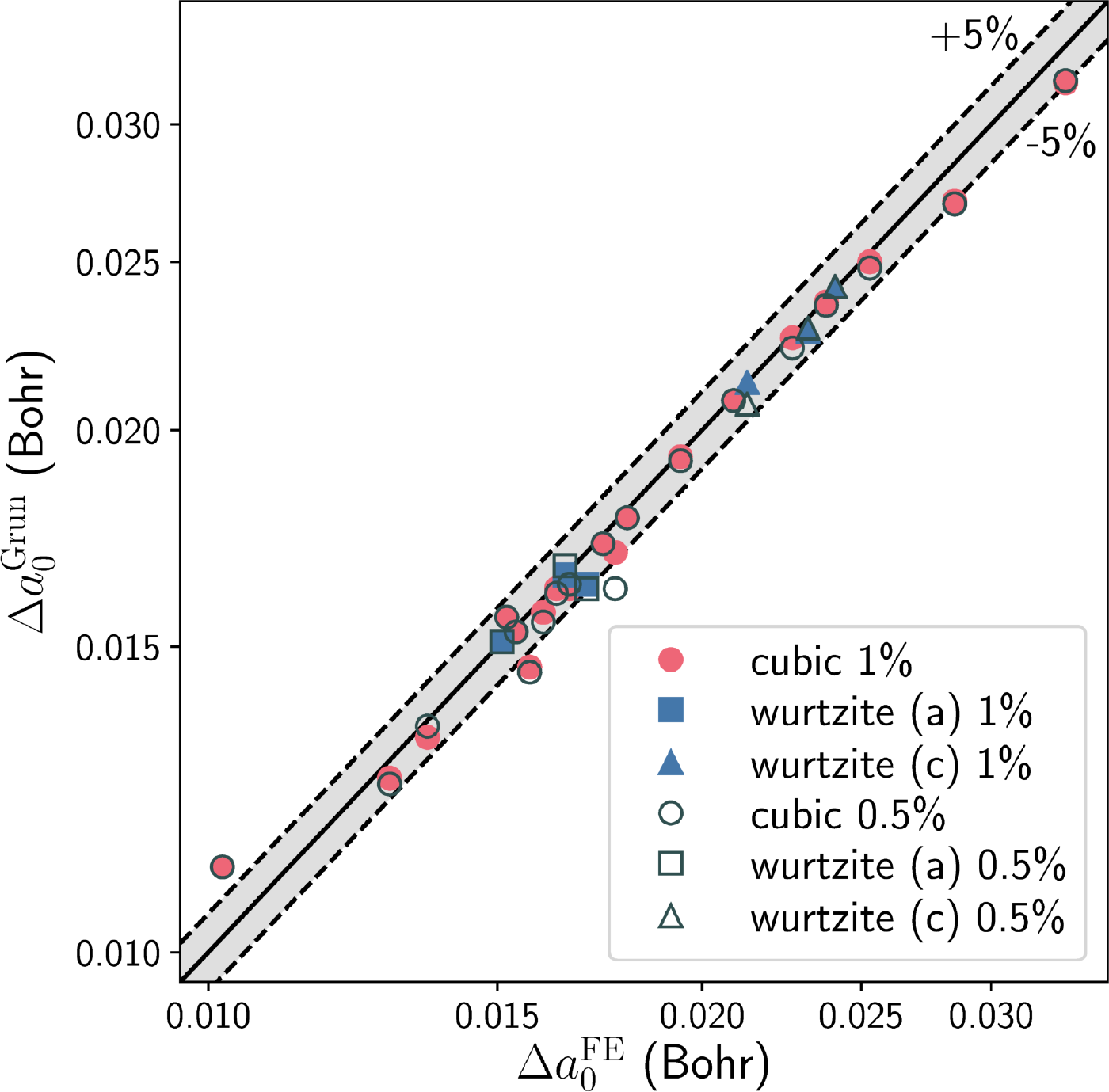}
    \caption{\textbf{Zero-point lattice expansion from FE minimization and \gru formalism.} The finite differences for mode \gru parameters were computed with a volumic variation of 1\% (solid colors) and 0.5\% (empty gray markers). Note the use of logarithmic scales. The shaded gray area corresponds to the region where both approaches agree within 5\% of each other. The ZPLE in the \gru approach tends to be slightly smaller than with the FE minimization. Most materials show a better agreement between both methodologies when using a 1\% variation. }
    \label{fig:grun_vs_fe_zplat}
\end{figure}
\begin{table*}[]
    \centering
    \caption{\textbf{Statistical errors on the ZPLE for the \gru approach.} The free energy minimization ZPLE results are taken as reference value. The mean absolute error (MAE), root-mean-square error (RMSE), and mean absolute relative error (MARE) for both the linear expansion and the resulting volumic expansion are displayed for the cubic materials, wurtzite materials and for the entire set.}
    \label{tab:zprlatt_stats}
      \setlength\extrarowheight{2pt}
    \begin{tabularx}{\textwidth}{p{26mm} *{6}{Y}}
    \hline\hline
       \multirow{3}{*}{Set}& \multicolumn{3}{c}{\gru 1\%} & \multicolumn{3}{c}{\gru 0.5\%}\\
         &   MAE  &  RMSE
       & MARE &   MAE  &  RMSE
       & MARE\\
       \cmidrule(lr){2-4}\cmidrule(lr){5-7}
       & ($10^{-3}$ Bohr) &($10^{-3}$ Bohr)&(\%)
       & ($10^{-3}$ Bohr) &($10^{-3}$ Bohr)&(\%)\\[2pt]
       \hline 
        Cubic~($a$)& 0.5 & 0.7 & 2.6 & 0.6 & 0.8 & 3.0\\
       Wurtzite~($a$) & 0.2 & 0.4 & 1.4 & 0.3 & 0.5 & 2.0\\
       Wurtzite~($c$) & 0.2 & 0.2 & 0.7 & 0.3 & 0.4 & 1.5\\
       Wurtzite~($a, c$)& 0.2 & 0.3 & 1.0 & 0.3 & 0.4 & 1.7\\
       \hline
        Full set~(linear) & 0.4 & 0.6 & 2.2 & 0.5 & 0.7 & 2.7\\
       \hline
              \multirow{3}{*}{Set}& \multicolumn{3}{c}{\gru 1\%} & \multicolumn{3}{c}{\gru 0.5\%}\\
        \cmidrule(lr){2-4}\cmidrule(lr){5-7}
         &   MAE  &  RMSE
       & MARE &   MAE  &  RMSE
       & MARE\\
       & (Bohr$^3$) &(Bohr$^3$)&(\%)
       & (Bohr$^3$) &(Bohr$^3$)&(\%)\\[2pt]
       \hline 
    Cubic~($V$)& 0.134 & 0.176 & 2.6 & 0.157 & 0.214 & 3.0 \\
      Wurtzite~($V$) & 0.030 & 0.041 & 1.3 & 0.032 & 0.044 & 1.4\\
       \hline
       Full set~($V$) & 0.120 & 0.164 & 2.4 & 0.140 & 0.199 & 2.8\\
       \hline\hline
       \end{tabularx}
       \end{table*}

From Tables~\ref{tab:a0data} and~\ref{tab:a0c0data}, we now focus on the predicted ZPLE from the \gru approach (Eq.~(\ref{eq:zpr-latt-grun}), \mbox{$\Delta a^{\textrm{Grun}} (T=0)$}. First, comparing results obtained with the 1\% and 0.5\% volume variations for the finite differences, we find that the discrepancy is at most 0.2~mBohr for most materials, the few exceptions being SiC, CdS and the $c$ parameter of ZnO. Thus, the materials in our dataset do not display great sensitivity to the configuration sampling used for the finite derivatives.

As the \gru formalism is by construction an approximation of the FE approach within the QHA framework, one should naturally compare the accuracy of this method to the outcome of FE minimization. At this point, we stress that our goal is to provide a numerical \enquote{proof of concept} of the equivalent predictive capabilities of these two methodologies. Hence, we solely compare the ZPLE values obtained within our numerical calculations. For reasons that will be clarified further in this article (see Sec.~\ref{sec:discussion}), assessing the respective predictability of the two approaches with respect to experimental data would benefit from a full temperature-dependent comparative analysis, which is beyond the scope of the current study.

We observe a good general agreement between the methodologies, as is emphasized by Fig.~\ref{fig:grun_vs_fe_zplat}. Note that the axes have a logarithmic scale. The relative error on the ZPLE with respect to FE minimization is under 5\% for all but 6 of our 22 materials (see the shaded gray area) and below 3\% for more than half of them. We stress that this error is relative to an already very small quantity; the relative error with respect to the renormalized lattice parameter, i.e. $a^0 + \Delta a\,(T=0)$, would be significantly smaller. The absolute errors for all materials are detailed in Tables~S2
and~S3
of the Supplemental Material~\cite{note_supplemental_material}. We also find that the agreement between both methodologies tends to decrease for materials with a larger bulk modulus, i.e. BN, MgO and C. The exceptions to this trend are CdS, as well as GaAs and Ge. For the two latter materials, we question the capability of the PBE functional to properly describe the electronic structure, as was suggested in Ref.~\cite{miglio_predominance_2020}. We nevertheless restrain our analysis to this functional for consistency with the other materials in our dataset.

Two other observations can be drawn from Tables~\ref{tab:a0data} and~\ref{tab:a0c0data}. On the one hand, we can see that the 1\% volumic variation for the \gru ZPLE improves the overall agreement with the FE ZPLE compared to the 0.5\% variation. This can be counter-intuitive as one typically aims for smaller variable sampling when approximating derivatives by finite differences. On the other hand, we notice that the \gru formalism tends to predict a slightly  smaller ZPLE compared to the FE minimization. The only two exceptions to this trend are Ge and GaP. In that case, this behavior could be attributed to our choice of the experimental lattice parameter as a static reference, which required us to add an artificial pressure to the FE in order to compute the lattice expansion, as explained in Sec.~\ref{sec:gsopt}.

The mean absolute error (MAE), root mean square error (RMSE) and mean absolute relative error (MARE)  for the \gru approach are shown in Table~\ref{tab:zprlatt_stats}, taking the FE results as a reference. Detailed results for the cubic and wurtzite materials are also provided. The statistical analysis for wurtzites ($a,c$) and the full set (linear) considers each independent lattice parameter as a single entry. Without surprise, the statistical errors decrease for the 1\% variation compared to the 0.5\% variation, in agreement with our previous observation. The cubic materials show the largest deviations in the set, followed respectively by the wurtzite $a$ and $c$ parameters. This could, however, be biased by the small proportion of wurtzite materials in our set compared to cubic ones. For the linear ZPLE ($\Delta a\,(T=0)$), we obtain a MAE of 0.4~mBohr, a RMSE of 0.6~mBohr and a MARE of 2.2\% for our entire dataset.  This yields in a MRE of 0.120~Bohr$^3$, RMSE of 0.164~Bohr$^3$ and MARE of 2.4\% for the zero-point volume expansion ($\Delta V\,(T=0)$). Should we exclude Ge and GaAs from the statistical analysis for the reason discussed earlier, the MARE significantly drops by half a percent in each case, to 1.7\% in the linear case and to 1.9\% in the volumic case.

To fully appreciate these results, one has to recall that the derivation of the \gru ZPLE relies on the lowest order approximation of the quasi-harmonic FE (see Sec.~\ref{sec:theo-gru}). In this approach, the ZPLE is described by the variation of the phonon frequencies with respect to the lattice parameters, truncated to first order. The absence of higher-order terms in the FE expansion is most likely responsible for the small discrepancies between both methods. Nevertheless, our results clearly show that the first-order \gru formalism can accurately account for the ZPLE, both for isotropic and anisotropic materials.

One should finally appreciate that, for low-symmetry materials, the FE can become up to a six-dimensional hypersurface if there are no symmetry constraints that prevent temperature-induced shear strains of the unit cell, not even counting the internal degrees of freedom, as detailed earlier. This will rapidly increase the minimal number of strained configurations that must be explicitly evaluated. While some promising methodologies have recently been developed to circumvent this difficulty~\cite{pike_calculation_2019, hellman_temperature_2013, huang_efficient_2016, zacharias_fully_2020}, a proper sampling of the FE surface can rapidly become a computational bottleneck.
As pointed out for the wurtzite materials, one could try circumventing the explicit sampling of the multidimensional FE by minimizing a volume-dependent FE curve and relaxing both the internal coordinates and
lattice parameters under a fixed volume constraint. Whether such an algorithm would correctly capture strong elastic anisotropy remains to be investigated. The \gru formalism, on the other hand, requires only $n+1$ configurations per degree of freedom, where $n$ is the desired accuracy level chosen for the central finite difference. Hence, the lower the symmetry, the greater the reduction of the numerical cost. For example, computing the ZPLE for the wurtzite materials, which have only two~degrees of freedom, required minimally more than 20~configurations for the 2D FE surface, while the \gru approach provided the same result from only five~configurations. Within our study we estimate the net gain to be about a factor of 3 for the cubic materials and a factor of at least 5 for the wurtzites.

\subsection{ZPLE renormalization of the band gap energy}\label{sec:res-gap}
\begin{table}[]
    \centering
    \caption{\textbf{Zero-point lattice expansion renormalization of the band gap.} The VBM and CBM locations of the fundamental band gap in the BZ are specified in the second column. The band gap renormalization was computed from static DFT using the ZPLE-renormalized lattice parameters obtained from FE minimization ($\Delta E_g^{\textrm{FE}}$) and with the \gru formalism ($\Delta E_g^{\textrm{Grun}}$), shown in Tables~\ref{tab:a0data} and~\ref{tab:a0c0data}.}
    \label{tab:ZPlatt-data}
\sisetup{
table-format = 5.2 ,
table-number-alignment = center ,
table-column-width = 1.65cm ,
}
     \setlength\extrarowheight{2pt}
    \begin{tabularx}{\columnwidth}{p{13mm} S S S S}
    \hline\hline
       \multirow{2}{*}{Material}   & {\multirow{2}{*}{Gap}} & {$\Delta E_g^{\textrm{FE}}$} & {$\Delta E_g^{\textrm{Grun 1\%}}$} & {$\Delta E_g^{\textrm{Grun 0.5\%}}$}\\
        & & {(meV)} & {(meV)} & {(meV)}\\[2pt]
       \hline
       C-dia & {$\Gamma$-$\Delta$}& -26.74  &  -26.48 & -26.23\\
       Si-dia  & {$\Gamma$-$\Delta$} & 8.71  & 8.60 &8.61\\
        Ge-dia &  {$\Gamma$-L} &-9.34 &  -10.23 &-10.24\\
        SiC-zb &  {$\Gamma$-X} &6.97  & 6.94 & 6.86\\
        \hline
        BN-zb & {$\Gamma$-X} & -16.89   & -16.07 & -16.02\\
        BAs & {$\Gamma$-$\Delta$}& 8.75  & 7.40 & 7.38\\
        AlP-zb & {$\Gamma$-X} & 9.74   & 9.60 & 9.60\\
        AlAs-zb & {$\Gamma$-X} & 7.40   & 7.24 & 7.27\\
        AlSb-zb &{$\Gamma$-L} & -11.33  &  -11.25 &  -11.22\\
        AlSb-zb &{$\Gamma$-$\Delta$}&  5.45 &  5.41 & 5.40\\
        GaN-zb & {$\Gamma$-$\Gamma$}& -47.49  & -47.22 & -47.25\\
        GaP-zb & { $\Gamma$-$\Delta$}& 7.60   & 7.79 & 7.79\\
        GaAs-zb & {$\Gamma$-$\Gamma$}& -31.37 &  -29.06 & -28.88\\
        \hline
        ZnS-zb & {$\Gamma$-$\Gamma$}& -24.23   & -24.11 & -23.99\\
        ZnSe-zb & {$\Gamma$-$\Gamma$}& -17.19  & -16.88 & -16.88\\
        ZnTe-zb & {$\Gamma$-$\Gamma$}& -17.90   &  -17.81 & -17.81\\
        CdS-zb & {$\Gamma$-$\Gamma$}& -10.50   & -10.07 & -9.62\\
        CdSe-zb & {$\Gamma$-$\Gamma$}& -7.17  & -6.83 & -7.12\\
        CdTe-zb & {$\Gamma$-$\Gamma$}& -9.03  &  -8.78 & -8.78\\
        MgO-rs & {$\Gamma$-$\Gamma$}& -116.16 & -110.75 &-111.17\\
        \hline
        AlN-w & {$\Gamma$-$\Gamma$}& -78.77 & -76.13 & -75.55 \\
        GaN-w & {$\Gamma$-$\Gamma$}& -49.19   & -48.42 & -48.49 \\
        ZnO-w & {$\Gamma$-$\Gamma$}& -11.43   &  -11.46 & -11.28 \\
          \hline\hline
    \end{tabularx}
\end{table}

\subsubsection{Free energy minimization vs \gru approach}
\begin{figure}
    \centering
    \includegraphics[width=\linewidth]{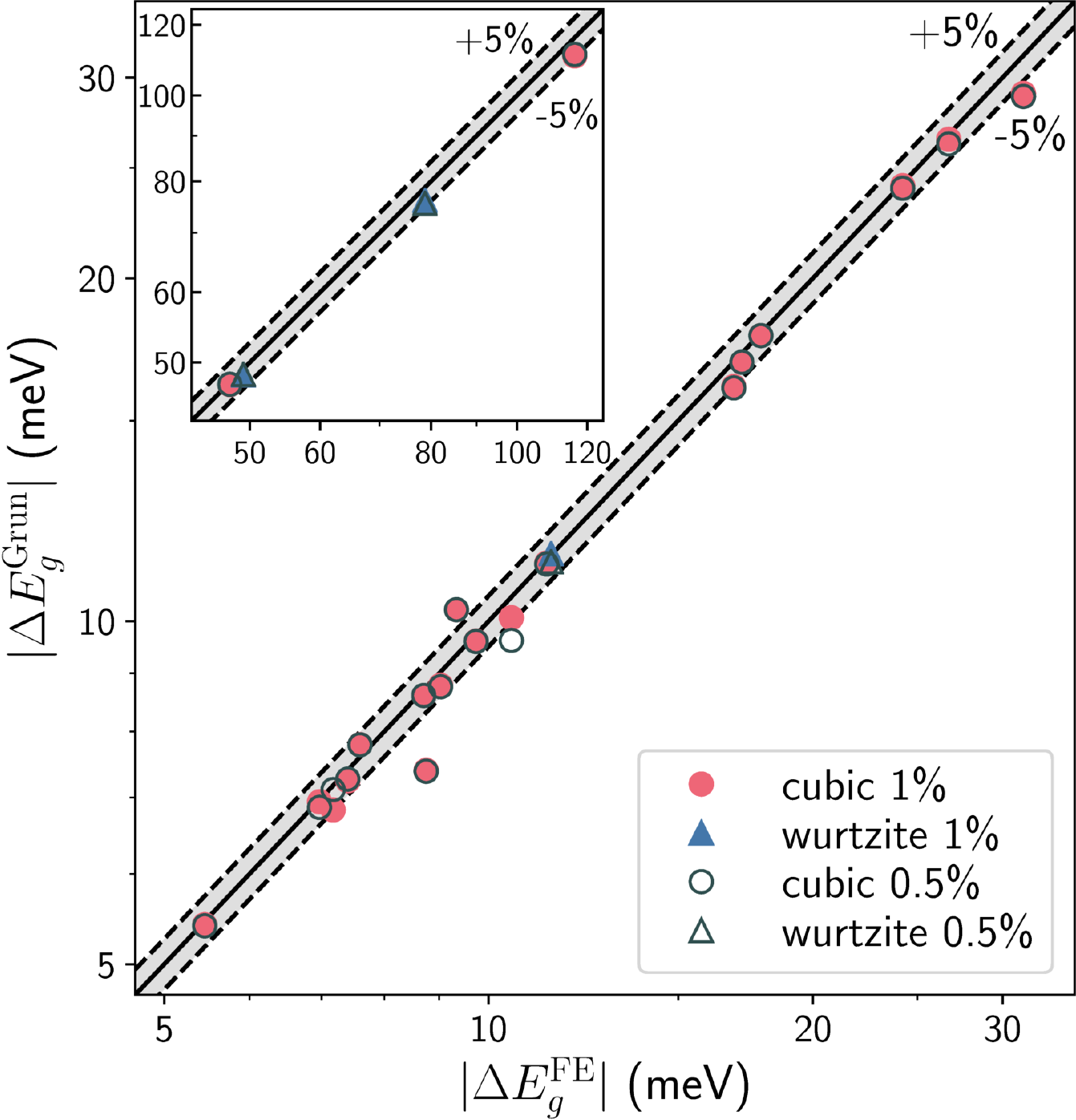}
    \caption{\textbf{Band gap ZPR induced by zero-point lattice expansion from FE minimisation and \gru formalism.} The finite differences for mode \gru parameters were computed with volumic variation of 1\% (solid colors) and 0.5\% (empty gray markers). Our results show that for most materials, the difference between the two sampling sets is below meV accuracy, hence has no significant impact on $\textrm{ZPR}_g^{\textrm{ZPLE}}$. Note the use of logarithmic scales. The shaded gray area corresponds to the region where both approaches agree within 5\% of each other. Materials whose $\textrm{ZPR}_g^{\textrm{ZPLE}}$ is over 40 meV are displayed in the inset to enhance clarity.}
    \label{fig:grun_vs_fe_zprg}
\end{figure}

The ZPLE contribution to the band gap ZPR, $\textrm{ZPR}_g^{\textrm{ZPLE}}$, is reported in Table~\ref{tab:ZPlatt-data}. To avoid any confusion with the full band gap ZPR or EPI-induced ZPR when we need to specify the calculation method, we denote it as $\Delta E_g$.  Note that for AlSb, the renormalization was computed for two conduction valleys: the $L$ point is the theoretical CBM extracted from the static DFT calculation, and the minimum along the \mbox{$\Delta = \Gamma-X$} high symmetry path corresponds to the experimental location of the fundamental band gap. From the two rightmost columns, we find that the choice of volume sampling for the finite differences used in the \gru approach shows a negligible impact on the final value of $\textrm{ZPR}_g^{\textrm{ZPLE}}$. The 1\% and 0.5\% volumic variation results agree with each other within meV accuracy, which is of the same order as the precision typically reported for ZPR corrections. This is shown on Fig.~\ref{fig:grun_vs_fe_zprg}, where the open and solid markers coincide almost perfectly. Note the axes' logarithmic scale again.

Making the comparison to FE results, we find again an agreement of about 1~meV, except for MgO ($\approx5$~meV) and AlN ($\approx3$~meV), as well as for GaAs ($\approx3$~meV). In the former case, these materials display by far the largest $\textrm{ZPR}_g^{\textrm{ZPLE}}$, such that the relative error remains coherent with the rest of the set. These three materials are also among those for which the absolute ZPLE discrepancy between FE and \gru was among the largest. Just as for the ZPLE, the \gru approach provides a very reliable value for $\textrm{ZPR}_g^{\textrm{ZPLE}}$, slightly smaller than the FE minimization. This can be understood as the numerical differences observed in Tables~\ref{tab:a0data} and~\ref{tab:a0c0data} between the two methods are in general on the order of the numerical precision typically reported in the literature for lattice parameters. Ge and GaP are again exceptions to this trend due to the small overestimation of the ZPLE by the \gru method.

If we rather focus on the sign of $\textrm{ZPR}_g^{\textrm{ZPLE}}$, we can notice some general trends for the different band gap locations in the BZ. Indeed, all materials displaying a direct fundamental band gap find their gap energy decreased by ZPLE. In indirect band gap materials, the gap energy also decreases for CBM located at the $L$ point, while it increases for gaps located along the $\Gamma-X$ high-symmetry path, diamond and BN being the only exceptions. To gain some insight into this trend, recall that increasing pressure, thus decreasing the lattice parameters, shifts the electronic bands to higher absolute energies due to the kinetic energy cost of bringing the ions closer to one another.
Since ZPLE is linked to lattice expansion, hence a negative external pressure, the sign of the band gap correction will therefore be determined by how relatively fast the VBM and CBM rise in energy with increased external pressure.  We thus recover the $-\partial E_g/\partial P$ dependency derived in Eq.~(\ref{eq:ZPLE-band_estimation}), yielding a negative $\textrm{ZPR}_g^{\textrm{ZPLE}}$ for direct band gaps at $\Gamma$.  For indirect band gaps, this pressure-induced energy shift can be partly or fully counteracted by the phase factor entering the Bloch states at finite $\kpoint$s.
The sign difference observed between diamond and silicon, which have the same lattice structure and whose conduction band edge lie on similar high symmetry lines, has been explained\cite{fahy_pressure_1987} in terms of the presence or absence of $d$ states with the same principal quantum number as the $p$-like valence states and the $s$-like conduction states, which drives a sign change in the pressure dependence of the band gap energy, hence, from Eq.~(37), in the ZPR$_g^{\rm{ZPLE}}$. In contrast with silicon (and with the zincblende materials whose CBM is at or close to the $X$ point), whose band edges have principal quantum number $n=3$, diamond (as well as zincblende BN) has $n=2$, for which no unoccupied $d$ states exist. See Table~VI for the values of $\partial E_g/\partial P$.

The statistical errors for $\textrm{ZPR}_g^{\textrm{ZPLE}}$ are displayed in Table~\ref{tab:zpgap_stats}. In agreement with the errors reported in Table~\ref{tab:zprlatt_stats} for the ZPLE, we find that the MARE is slightly larger for the cubic materials compared to the wurtzites. However, the larger MAE and RMSE observed for the wurtzites can be attributed to their larger ZPR values compared to the cubic set. For the entire dataset, we obtain a MAE of 0.74~meV, RMSE of 1.42~meV and MARE of 3.1\%, thus confirming the reliability of the \gru method for predicting $\textrm{ZPR}_g^{\textrm{ZPLE}}$ within the QHA.

\begin{table*}[]
    \centering
    \caption{\textbf{Statistical errors on the zero-point lattice expansion renormalization of the band gap.} The FE minimization results are taken as reference value ($\Delta E_g^{\textrm{FE}}$ from Table~\ref{tab:ZPlatt-data}). The \gru approximation with resp. $\pm$1\% and $\pm$0.5\% volumic variation were used. The mean absolute error (MAE), root-mean-square error (RMSE), and mean absolute relative error (MARE) are displayed for the cubic materials, wurtzite materials and for the entire set.}
    \label{tab:zpgap_stats}
      \setlength\extrarowheight{2pt}
    \begin{tabularx}{\textwidth}{p{20mm} *{6}{Y}}
    \hline\hline
       \multirow{3}{*}{Set}& \multicolumn{3}{c}{\gru 1\%} & \multicolumn{3}{c}{\gru 0.5\%}\\
       \cmidrule(lr){2-4}\cmidrule(lr){5-7}
                &   MAE  &  RMSE
       & MARE &   MAE  &  RMSE
       & MARE\\
       & (meV) &(meV)&(\%) & (meV) &(meV)&(\%)\\[2pt]
       \hline 
       Cubic & 0.68 & 1.39 & 3.3 & 0.70 & 1.34 & 3.5\\
       Wurtzite & 1.14 & 1.59 & 1.7 & 1.36 & 1.91 & 2.3\\
       \hline
       Full set & 0.74 & 1.42 & 3.1 & 0.79 & 1.43 & 3.4\\
       \hline\hline
       \end{tabularx}
       \end{table*}
       
\subsubsection{Comparison to the  empirical expression}\label{sec:res-estimate}
\begin{table*}[]
    \centering
    \caption{\textbf{Estimation of the zero-point lattice expansion renormalization of the band gap,} using Eq.~(\ref{eq:ZPLE-band_estimation}). The band gap edges, bulk modulus extracted from elastic constants and pressure derivative of the band gap used to compute the estimates are provided for the different materials. The $\Delta V\,(T=0)/V^0$ term was evaluated in three different ways: from the numerical data presented in Tables~\ref{tab:a0data} and~\ref{tab:a0c0data} for 1\% \gru parameters ($\Delta E_g^{\textrm{data}}$), from the extrapolation to $T=0$~K of the high-$T$ slope of our theoretical calculation of $\Delta a\,(T)/a^0$ obtained from Eq.~(\ref{eq:alpha-strain}) and~(\ref{eq:aoft-strain}) using 1\% \gru parameters, and from the same extrapolation technique applied to available experimental data ($\Delta E_g^{\textrm{exp}}$, see text for details). Two estimations are presented using the theoretical high-$T$ slope method: the first takes the slope between $T\approx2\theta_D$ and $T\approx3\theta_D$ (except for diamond, SiC and BN, see text) ($\Delta E_g^{T>2\theta_D}$), while the second takes the slope between 700 and 1000~K ($\Delta E_g^{700-1000{\textrm{K}}}$), regardless of the Debye temperature. This emphasizes that the accuracy of the estimation strongly relies on the choice of the \enquote{high-$T$ regime}. $\textrm{ZPR}_g^{\textrm{Grun 1\%}}$ (from Table~\ref{tab:ZPlatt-data}) is taken as reference to assess the accuracy of the estimation. The numerical values of $\Delta a_k\,(T=0)/a_k^0$ used to obtain the different estimates are explicited in Table~S4
    of the Supplemental Material~\cite{note_supplemental_material}.}
    \label{tab:zpgap-estimation}
\sisetup{
table-format = 5.2 ,
table-number-alignment = center ,
table-column-width = 1.6cm ,
}
     \setlength\extrarowheight{2pt}
    \begin{tabularx}{\textwidth}{p{14mm} S S[table-format=4.0] S[table-format=3.0] S[table-format=4.1] S S S S S}
    \hline\hline
       \multirow{2}{*}{Material}   & {\multirow{2}{*}{Gap}} & {$\theta_D$~\cite{madelung_semiconductors_2004}} & {$B_0^{\textrm{elast}}$} & {$\partial E_g/\partial P$} &{$\Delta E_g^{\textrm{Grun 1\%}}$} & {$\Delta E_g^{\textrm{\textrm{data}}}$} & {$\Delta E_g^{T>2\theta_D}$} & {$\Delta E_g^{700-1000{\textrm{K}}}$}&  {$\Delta E_g^{\textrm{exp}}$}\\
       & & {(K)}& {(GPa)} & {(meV/GPa)} &{(meV)} & {(meV)} & {(meV)} & {(meV)} & {(meV)}\\[2pt]
       \hline
       C-dia & {$\Gamma$-$\Delta$}& 2220& 431 & 5.5 & -26.48 & -26.15& -22.39 &-11.78& -27.06~\cite{cardona_isotope_2005}\\
       Si-dia  & {$\Gamma$-$\Delta$} &636& 88 & -19.7 & 8.61 &8.67 & 7.43 & 6.43   & 9.72~\cite{cardona_isotope_2005}\\
        Ge-dia &  {$\Gamma$-L} &374& 87 & 34.1 & -10.24 &  -9.44 & -8.04 &-7.88  & -12.76~\cite{cardona_isotope_2005}\\
        SiC-zb &  {$\Gamma$-X} &1270 &  211 & -4.0 & 6.94  & 7.01 & 6.14&4.24 & 6.95~\cite{reeber_thermal_1995}\\
        \hline
        BN-zb & {$\Gamma$-X} & 1730 &403 & 3.3& -16.07   & -15.81 & - 13.23 &-7.84\\
        BAs & {$\Gamma$-$\Delta$}& 800 &131 & -7.2 & 7.40  & 7.35 & 6.26 &4.95  & 3.56~\cite{kang_basic_2019}\\
        AlP-zb & {$\Gamma$-X} & 588 &82 & -22.9 & 9.60   & 9.69 & 8.36 &7.39\\
        AlAs-zb & {$\Gamma$-X} & 414 &67 & -24.2 & 7.24   & 7.33 & 6.20& 5.95 \\
        AlSb-zb &{$\Gamma$-L} & 292 &49 & 55.5 & -11.25  &  -11.29 &  -9.14 &-9.42\\
        AlSb-zb &{$\Gamma$-$\Delta$}& 292& 49 & -26.8& 5.42 &  5.44 & 4.41& 4.55\\
        GaN-zb & {$\Gamma$-$\Gamma$}& 608 &172 & 37.9 & -47.22  & -47.34 & -38.90&-32.91\\
        GaP-zb & { $\Gamma$-$\Delta$}& 445 &86 & -25.5 & 7.79   & 9.98 & 8.46 & 8.00  &11.40~\cite{reeber_thermal_1995}\\
        GaAs-zb & {$\Gamma$-$\Gamma$}& 344&61 & 118.8 & -29.06 &  -29.16 & -25.48 &-25.30 &\\
        \hline
        ZnS-zb & {$\Gamma$-$\Gamma$}& 352 &70 & 61.7 & -24.11   & -24.22 & -20.48&-20.31\\
        ZnSe-zb & {$\Gamma$-$\Gamma$}& 339 &57 & 68.7 & -16.88  & -16.93 & 15.04& -14.98\\
        ZnTe-zb & {$\Gamma$-$\Gamma$}& 180 &43 & 105.0 & -17.81   &  -17.78 &-14.75 &-16.08\\
        CdS-zb & {$\Gamma$-$\Gamma$}& 219 &53 & 41.6 & -10.07   & -10.06 & -7.89&-8.57\\
        CdSe-zb & {$\Gamma$-$\Gamma$}& 182 &45 & 46.3 & -6.83  & -7.05 & -5.86 &-6.35\\
        CdTe-zb & {$\Gamma$-$\Gamma$}& 158& 35 & 83.5 & -8.78  &  -8.83 & -7.36& -8.12\\
        MgO-rs & {$\Gamma$-$\Gamma$}& 948 &149 & 63.5 &  -110.75 & -112.11 & -102.50&-86.49  & -110.60~\cite{dubrovinsky_thermal_1997, fiquet_high-temperature_1999}\\
        \hline
        AlN-w & {$\Gamma$-$\Gamma$}&950 &208 &42.4 &-76.13 & -72.87 & -63.02 & -46.34& -70.44~\cite{wang_thermal_1997}\\
        GaN-w & {$\Gamma$-$\Gamma$}&600 &172 &39.2 &-48.42   & -49.04 & -37.77 & -34.11 & -35.73~\cite{roder_temperature_2005}\\
        ZnO-w & {$\Gamma$-$\Gamma$}& 440 &130 & 12.6& -11.46   & -12.25 & -9.88 & -9.22  & -15.80~\cite{reeber_lattice_1970,iwanaga_anisotropic_2000}\\
          \hline\hline
    \end{tabularx}
\end{table*}
\begin{table*}[]
    \centering
    \caption{\textbf{Statistical errors on the estimation of the ZPLE-induced band gap ZPR from Eq.~(\ref{eq:ZPLE-band_estimation}),} using the three different methods explained in the caption of Table~\ref{tab:zpgap-estimation}. For consistency, the ZPLE-induced band gap ZPR obtained with 1\% \gru parameters are taken as reference value ($\Delta E_g^{\textrm{Grun 1\%}}$ from Table~\ref{tab:ZPlatt-data}, see text). The mean absolute error (MAE), root-mean-square error (RMSE), and mean absolute relative error (MARE) are displayed for the cubic materials, wurtzite materials and for the entire set. $\Delta E_g^{\textrm{\textrm{data}}}$ quantifies the accuracy of the estimation obtained from Eq.~(\ref{eq:ZPLE-band_estimation}), while $\Delta E_g^{T>2\theta_D}$ and $\Delta E_g^{700-1000{\textrm{K}}}$ reveal its sensitivity to the choice of temperature range used for the extrapolation of $\Delta a\,(T=0)/a^0$.}
    \label{tab:zpgap_est_stats}
      \setlength\extrarowheight{2pt}
    \begin{tabularx}{\textwidth}{p{20mm} *{9}{Y}}
    \hline\hline
       \multirow{3}{*}{Set}& \multicolumn{3}{c}{$\Delta E_g^{\textrm{\textrm{data}}}$} & \multicolumn{3}{c}{$\Delta E_g^{T>2\theta_D}$} &
       \multicolumn{3}{c}{$\Delta E_g^{700-1000{\textrm{K}}}$}\\
       \cmidrule(lr){2-4}\cmidrule(lr){5-7}\cmidrule(lr){8-10}
       &   MAE  &  RMSE
       & MARE &   MAE  &  RMSE
       & MARE&   MAE  &  RMSE
       & MARE\\
       & (meV) &(meV)&(\%) & (meV) &(meV)&(\%)& (meV) &(meV)&(\%)\\[2pt]
       \hline 
       Cubic & 0.30 & 0.62 & 2.5 & 2.58 & 3.36 & 15.1 & {~}4.57 & {~}7.58 & 21.7\\
       Wurtzite & 1.56 & 1.97 & 4.2 & 8.45 & 9.80 & 16.7 & 15.45 & 19.13 & 29.4\\
       \hline
       Full set & 0.47 & 0.91 & 2.7 & 3.34 & 4.72 & 15.4 & {~}5.99 & {~}9.88 & 22.7\\
       \hline\hline
       \end{tabularx}
       \end{table*}
\begin{figure}
    \centering
      \includegraphics[width=\linewidth]{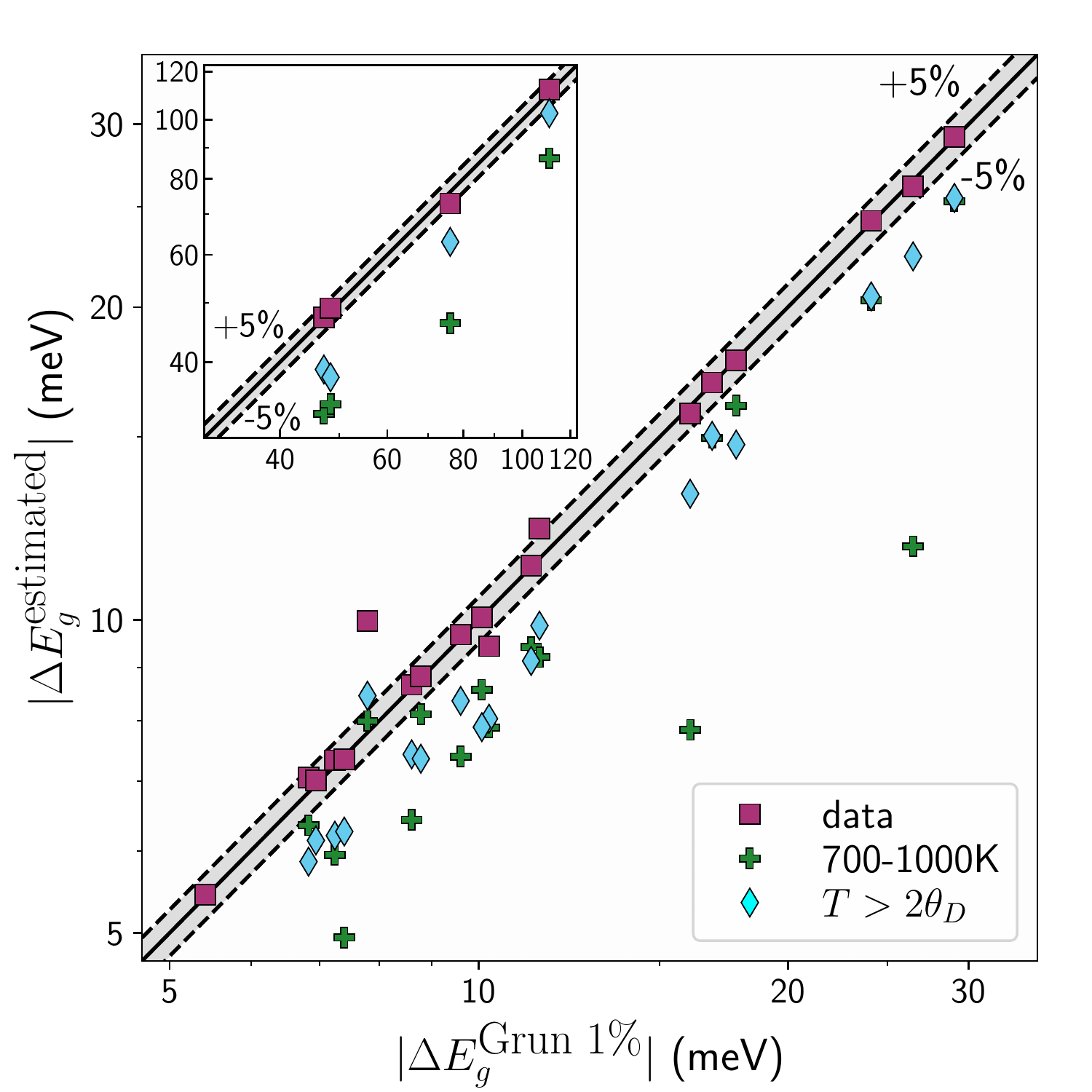}
    \caption{\textbf{Estimation of the ZPLE-induced band gap renormalization from the empirical expression} Eq.~(\ref{eq:ZPLE-band_estimation}), using $\Delta a\,(T=0)/a^0$ as computed from 1\% \gru parameters (purple squares), and extracted from the $T=0$~K intercept of a linear fit $\Delta a\,(T)/a^0$ in the high temperature regime with $T\in[700, 1000]$~K (green plus) and $T\in[2\theta_D, 3\theta_D]$ (cyan diamonds). Note that the scales are logarithmic. Numerical results from Tables~\ref{tab:a0data} and~\ref{tab:a0c0data} are taken as reference values.}
    \label{fig:estimation}
\end{figure}

To assess the accuracy of the empirical expression of Eq.~(\ref{eq:ZPLE-band_estimation}), we evaluate $\Delta V/V^0\,(T=0~\text{K})$ in 3 different ways. First, we make use of the numerical results presented in Tables~\ref{tab:a0data} and~\ref{tab:a0c0data} ($\Delta E_g^{\textrm{\textrm{data}}}$) for 1\% \gru parameters. Second, we theoretically evaluate $\Delta a\,(T)/a^0$ from Eq.~(\ref{eq:alpha-strain}) and~(\ref{eq:aoft-strain}) using 1\% \gru parameters, and compute the intersect at $T=0$~K of a linear fit in the high temperature regime. We favor the \gru approach for this purpose as it delivers a smooth temperature dependence for the lattice parameters up to arbitrary high temperatures without the need for additional calculations, in contrast with the FE minimization method, for which our current volume sampling does not explicitly capture the FE minima at very high temperatures.
The third method evaluates the $T=0$~K intersect of the high-$T$ slope from available experimental data ($\Delta E_g^{\textrm{exp}}$). Details about our treatment of experimental data can be found in Sec.~S4
of the Supplemental Material~\cite{note_supplemental_material}. We finally investigate the sensitivity of the estimation to the choice of temperature regime for the linear fit by considering two different ranges for the second method. The first one is material-specific and covers $T\in[2\theta_D, 3\theta_D]$ ($\Delta E_g^{T>2\theta_D}$), while the second covers $T\in[700, 1000]$~K regardless of the Debye temperature ($\Delta E_g^{700-1000K}$). For diamond, Si and BN, we set the first range to be respectively $T\in[1.5\theta_D, 1.8\theta_D]$, $T\in[2\theta_D, 2.4\theta_D]$ and $T\in[1.5\theta_D, 1.8\theta_D]$, as otherwise, it would go beyond the melting temperature.

The estimation of $\textrm{ZPR}_g^{\textrm{ZPLE}}$ from the different methods, as well as relevant parameters used to obtain them, are presented in Table~\ref{tab:zpgap-estimation} and Fig.~{\ref{fig:estimation}}. The estimates are compared with the DFT-computed $\Delta E_g^{\textrm{Grun 1\%}}$ previously reported in Table~\ref{tab:ZPlatt-data}. Statistical errors are presented in Table~\ref{tab:zpgap_est_stats}. Note that all estimations rely on the same values of $B_0$ and $\partial E_g/\partial P$; hence, the numerical differences are solely explained by variations of $\Delta a/a^0$. Without surprise, $\Delta E_g^{\textrm{data}}$~(purple squares) provides the most reliable estimate, as it makes use of the same value of $\Delta a/a^0$ computed with the first-principles approach as the reference $\Delta E_g^{\textrm{Grun 1\%}}$, without relying on any extrapolation procedure. These results are thus a direct assessment of the accuracy of Eq.~(\ref{eq:ZPLE-band_estimation}). For the cubic materials, most materials display a remarkable numerical agreement below 0.5~meV (except, again,  GaP, Ge and MgO, for which the agreement is still nevertheless excellent), yielding a MARE of 2.5\%. The wurtzites materials show a slightly higher MARE of 4.2\%, which can most likely be attributed to AlN. In all cases, we obtain a global MARE of 2.7\% for the whole dataset, thus validating a posteriori Eq.~(\ref{eq:ZPLE-band_estimation}).

We now shift our focus on $\Delta E_g^{{T>2\theta_D}}$ (cyan diamonds) and $\Delta E_g^{\textrm{700-1000K}}$ (green plus), which both rely on the extrapolation of the linearized high-$T$ solution to \mbox{$T=0$~K} from our first-principles \gru results. In contrary to the previous assessment, both results display important discrepancies compared to the reference values. The statistical errors increase accordingly, yielding a MARE of resp. 15.4\% and 22.7\%. We can group the
general trends into three categories. First, for materials with a high Debye temperature (\mbox{$\theta_D\gtrsim 500$~K}),  choosing a fixed, arbitrary temperature range (\mbox{$\Delta E_g^{\textrm{700-1000K}}$}) strongly underestimates the ZPLE-induced band gap ZPR, while the material-dependent approach (\mbox{$\Delta E_g^{{T>2\theta_D}}$}) visibly improves the accuracy of the estimation. The difference is even more striking for materials with \mbox{$\theta_D>1000$~K}. In that case, the chosen fixed range does not cover a region where the classical regime approximation is physically valid, hence its incapacity to deliver a reliable estimate. A second category regroups materials whose Debye temperature is below room temperature. In
that case, the fixed 700-1000~K range delivers a better estimate than the material-specific approach. This observation may seem surprising at first,  but can be understood by recalling that the linearized behavior in the high-$T$ regime originates by construction from a first-order approximation of the classical Maxwell-Boltzmann distribution function, which may not be entirely accurate at intermediate temperatures (below 700~K), despite the material being in principle properly described by the classical picture. The third category regroups the few materials with \mbox{$\theta_D \in [300, 500]$~K}, for which the apparent agreement between the two methods is simply a consequence of their respective temperature range being almost identical.
The rightmost column of Table~\ref{tab:zpgap-estimation} computes the estimation of $\textrm{ZPR}_g^{\textrm{ZPLE}}$ from available experimental data. From our previous observations, we stress that such an analysis requires the lattice parameter to be measured in a broad range of temperature, namely, both sufficiently above the Debye temperature, to ensure that the linear fit is indeed in the true high-$T$ regime, and at low enough temperatures, to evaluate the experimental $T=0$~K value correctly. Our results nevertheless allow us to conclude that, when the experimental data range is appropriate for this type of analysis, Eq.~(\ref{eq:ZPLE-band_estimation}) provides a more than reliable reference for $\textrm{ZPR}_g^{\textrm{ZPLE}}$. In most cases, our numerical results are in very good agreement with the experimental estimation. They always minimally show the correct order of magnitude, thus validating our first-principles approach for computing the ZPLE correction to the band gap.

\subsubsection{Comparison of the ZPLE and EPI contributions to the band gap ZPR}

\begin{table}[]
    \centering
    \caption{\textbf{Zero-point renormalization of the band gap.} The ZPLE contribution to the band gap from FE minimization (Table~\ref{tab:ZPlatt-data}) is rounded to 1~meV. The rightmost column gives the relative variation of the band gap ZPR when considering the combined effects of ZPLE and EPI compared to EPI contribution only: $\textrm{R}^{\textrm{ZPLE/EPI}} = \textrm{ZPR}_g^{\textrm{ZPLE}}/\textrm{ZPR}_g^{\textrm{EPI}}=(\textrm{ZPR}_g^{\textrm{tot}}-\textrm{ZPR}_g^{\textrm{EPI}})/\textrm{ZPR}_g^{\textrm{EPI}}$. As $\textrm{ZPR}_g^{\textrm{EPI}}$ is negative, i.e. decrease of the band gap energy, a positive variation indicates that the total ZPR has a larger magnitude, hence is \enquote{more negative}, while a negative variation indicates a decrease in magnitude, as it becomes \enquote{less negative}.}
    \label{tab:ZPdata}
    \sisetup{
table-number-alignment = center ,
table-column-width = 1.55cm ,
}
     \setlength\extrarowheight{2pt}
    \begin{tabularx}{\columnwidth}{p{18mm} S[table-format=5.0] S[table-format=5.0] S[table-format=5.0] S[table-format=4.1]}
    \hline\hline
       \multirow{2}{*}{Material}  &  {$\textrm{ZPR}_g^{\text{EPI}}$} & {$\textrm{ZPR}_g^{\text{FE}}$} & {$\textrm{ZPR}_g^{\text{tot}}$}  & {$ \textrm{R}^{\textrm{ZPLE/EPI}}$}\\
        & {(meV)~\cite{miglio_predominance_2020}} & {(meV)} &{(meV)} & {(\%)} \\[2pt]
       \hline
       C-dia & -330 & -27 & -357 &  {+}8.2 \\
       Si-dia & -56 & 9 & -47  & -16.1 \\
        Ge-dia & -33& -9 & -42 &   {+}27.3\\
    SiC-zb & -179 & 7 & -172 &   -3.9 \\
        \hline
    BN-zb & -406 & -17 & -423  & {+}4.2\\
        BAs & -97 & 9 &-88 & -9.3\\
        AlP-zb & -93 & 10 & -83 &  -10.8\\
        AlAs-zb & -74 & 7 & -67 &  -9.5 \\
        AlSb-zb (L) & -43 & -11 & -54 &  {+}25.6 \\
        AlSb-zb ($\Delta$) & -51 & 5 & -46 &  -9.8 \\
        GaN-zb & -176 & -47 & -223  & {+}26.7\\
        GaP-zb & -65 & 8 & -57 & -12.3 \\
       GaAs-zb & -21 & -31 & -52 &  {+}147.6 \\
        \hline
        ZnS-zb & -88 & -24 & -112 &  {+}27.3\\
        ZnSe-zb & -44 & -17 & -61 &  {+}38.6 \\
        ZnTe-zb & -22 & -18 & -40 &  {+}81.8 \\
        CdS-zb & -70 & -10 & -80 &  {+}14.3 \\
        CdSe-zb & -34 & -7 & -41 &  {+}20.6 \\
        CdTe-zb & -20 & -9 & -29 &  {+}45.0 \\
    MgO-rs & -524 & -116 & -640  & {+}22.1\\
    \hline
    AlN-w & -399 & -79 & -478  & {+}19.8 \\
    GaN-w & -189 & -49 & -238 & {+}25.9 \\
    ZnO-w & -157 & -11 & -168 &  {+}7.0 \\
    \hline\hline
    \end{tabularx}
\end{table}
\begin{table*}[]
    \centering
    \caption{\textbf{Effect of exchange-correlation functional on the zero-point lattice expansion and zero-point renormalization of the band gap.} The fractional ZPLE, as well as the EPI and ZPLE contributions to the band gap ZPR, are compared for three semilocal XC functionals. The ratios $\textrm{R}^{\textrm{ZPLE/EPI}}$ give the relative variation of the band gap ZPR when considering the combined effects of ZPLE and EPI compared to EPI contribution only. See the caption of Table~\ref{tab:ZPdata} for the significance of the positive or negative sign. Despite their respective static equilibrium lattice parameters varying by more than the typical ZPLE correction, all three XC functionals yield ratios in good agreement with each other, thus validating the ratios presented in Table~\ref{tab:ZPdata} for all materials.
    }
    \label{tab:ZPdata_functionals}
    \sisetup{
table-number-alignment = center ,
table-column-width = 1.65cm ,
}
     \setlength\extrarowheight{2pt}
    \begin{tabularx}{\textwidth}{p{18mm} *{3}{S[table-format=2.4]}
    *{3}{S[table-format=1.3]}
    *{3}{S[table-format=5.0]} 
    }
    \hline\hline
       \multirow{2}{*}{Material}  & \multicolumn{3}{c}{$a^0$ (Bohr)} &
       \multicolumn{3}{c}{$\Delta a^{\textrm{FE}}/a^0$ (\%)} & \multicolumn{3}{c}{$\textrm{ZPR}_g^{\text{FE}}$ (meV)} \\
       \cmidrule(lr){2-4} \cmidrule(lr){5-7} \cmidrule(lr){8-10}
        & {PBE} & {PBESOL} &{LDA} & {PBE} & {PBESOL} &{LDA} & {PBE} & {PBESOL} &{LDA} \\[2pt]
       \hline
        GaAs-zb & 10.8627 & 10.6994 & 10.5861  &  0.145 & 0.127 & 0.128 & -31 & -28 & -29 \\
        ZnS-zb & 10.2859 & 10.1196& 10.0064& 0.189 & 0.180 & 0.176 & -24 & -23 & -24\\
        ZnSe-zb & 10.8330 & 10.6437& 10.5192& 0.148 & 0.136& 0.137&-17 & -16 & -17\\
        ZnTe-zb & 11.6819& 11.4675& 11.3316& 0.132 & 0.122 & 0.121&  -18 & -17 & -18  \\
    MgO-rs & 8.0366 & 7.9629& 7.8615& 0.414 &  0.412 & 0.394& -116 & -118  & -120\\
   \hline
    \end{tabularx}
    \begin{tabularx}{\textwidth}{p{18mm} 
    *{3}{S[table-format=5.0]} *{3}{S[table-format=5.0]} *{3}{S[table-format=5.1]}
    }
    \multirow{2}{*}{Material}  &\multicolumn{3}{c}{$\textrm{ZPR}_g^{\text{EPI}}$ (meV)} & \multicolumn{3}{c}{$\textrm{ZPR}_g^{\text{tot}}$ (meV)}  & \multicolumn{3}{c}{$ \textrm{R}^{\textrm{ZPLE/EPI}}$ (\%)}\\
    \cmidrule(lr){2-4} \cmidrule(lr){5-7} \cmidrule(lr){8-10}
    &{PBE} & {PBESOL} &{LDA} & {PBE} & {PBESOL} &{LDA} & {PBE} & {PBESOL} &{LDA} \\[-8pt]
    \hline
    GaAs-zb & -21 & -24 & -24 & -52 & -52& -53& {+}147.6 & {+}116.7& {+}120.8\\
    ZnS-zb & -88 & -82 & -78 & -112 & -105 & -102& {+}27.3 & {+}28.0 & {+}30.1\\
    ZnSe-zb & -44 &-41 &-40 & -61 & -57& -57& {+}38.6 & {+}39.0 & {+}42.5 \\
    ZnTe-zb & -22 & -21 & -22& -40 & -38 & -40& {+}81.8  & {+}81.0  & {+}81.8 \\
    MgO-rs & -524 & -511 & -526& -640 & -629 & -646& {+}22.1 & {+}23.1 & {+}22.8\\
    \hline\hline
    \end{tabularx}
\end{table*}

We finally compare the respective contributions of ZPLE and EPI to the total band gap ZPR in Table~\ref{tab:ZPdata}. Throughout the literature, the EPI contribution is commonly recognized as the leading mechanism driving the zero-point renormalization of the electronic bands, especially for materials containing light atoms, like B, C, N and O. To assess the importance of the ZPLE contribution, we also evaluate the relative variation of $\textrm{ZPR}_g$ in the full QHA  picture, with respect to EPI contribution only: $\textrm{R}^{\textrm{ZPLE/EPI}} = (\textrm{ZPR}_g^{\textrm{tot}} - \textrm{ZPR}_g^{\textrm{EPI}})/\textrm{ZPR}_g^{\textrm{EPI}} = \textrm{ZPR}_g^{\textrm{ZPLE}}/\textrm{ZPR}_g^{\textrm{EPI}}$. In this section, we only consider the FE methodology for the ZPLE contribution. As the results presented in Sec.~\ref{sec:res-gap} showed an excellent agreement between the FE and \gru approaches, the following observations will remain qualitatively valid should one rather consider the \gru ZPLE.

From the second leftmost column, we can see that EPI interaction reduces the band gap energy for all the materials considered. This is a typical behavior for semiconductors, known as the Varshni effect~\cite{varshni_temperature_1967}. 
As discussed earlier, the ZPLE contribution either closes or opens the band gap, depending on the band edge location in the BZ. As the magnitude of the $\textrm{ZPR}_g^{\textrm{ZPLE}}$ is smaller than its EPI counterpart, the resulting total band gap ZPR remains negative for all materials. The rightmost column shows the relative variation of the band gap ZPR induced by ZPLE. Note that since the total band gap ZPR is negative, a positive variation indicates that the EPI and ZPLE contributions reinforce each other; namely, the ZPR becomes more negative. A negative variation consequently signifies that the ZPLE contribution competes with EPI and thus reduces its impact on the total band gap ZPR.

If we rather consider the magnitude of the relative variation, $R^{\textrm{ZPLE/EPI}}$, we find that, out of our 22 materials, 7 of them see their total $\textrm{ZPR}_g$ modified by up to 10\% by the inclusion of ZPLE (C, AlSb, AlAs, SiC, BN, BAs and ZnO), 5 are modified by 10-20\% (Si, GaP, AlP, CdS and AlN), and 10 of them, hence almost half of our set, see their ZPR affected by 20\% or more (Ge, GaAs, CdTe, CdSe, ZnTe, ZnSe, ZnS, cubic and wurtzite GaN and MgO). For the two tellurides, which are the heaviest materials in the set, ZPR$_g^{\textrm{ZPLE}}$ is about half the size of ZPR$_g^{\textrm{EPI}}$ or more, which is all but negligible. Moreover, three materials containing N and O see their $\textrm{ZPR}_g$ increased by 20\%-25\%, which cautions about disregarding ZPLE effects even in materials containing light atoms.

We now briefly analyze the data in Table~\ref{tab:ZPdata} in terms of the elastic response of the materials ($B_0$) and their degree of ionicity. For this purpose, the three topmost sections of Table~\ref{tab:ZPdata} respectively separate group \MakeUppercase{\romannumeral 4}, \MakeUppercase{\romannumeral 3}-\MakeUppercase{\romannumeral 5} and \MakeUppercase{\romannumeral 2}-\MakeUppercase{\romannumeral 6} cubic materials. 
While no obvious quantitative trends can be extracted, we can nevertheless gain some qualitative insight into the relative strength of $\textrm{ZPR}_g^{\textrm{ZPLE}}$ for the different types of materials. First, despite the three exceptions mentioned earlier, materials with light atoms and high $B_0$ tend to be less affected by the inclusion of ZPLE, most likely thanks to their very large $\textrm{ZPR}_g^{\textrm{EPI}}$ (C, BN, SiC). As a corollary, heavier materials typically show smaller values of $\textrm{ZPR}_g^{\textrm{EPI}}$, hence are more strongly affected by $\textrm{ZPR}_g^{\textrm{ZPLE}}$ (see, for example, the Zn$X$ and Cd$X$ families, where $X$=\{S, Se, Te\}). Second, a higher degree of ionicity seems to be associated with a larger impact of $\textrm{ZPR}_g^{\textrm{ZPLE}}$ on the total band gap ZPR (i.e. \MakeUppercase{\romannumeral 2}-\MakeUppercase{\romannumeral 6} materials compared to  \MakeUppercase{\romannumeral 3}-\MakeUppercase{\romannumeral 5} materials). There are, however, outliers to both of these observations. Hence, a more thorough analysis of the effect of ZPLE on the EPI might prove useful to explain our results in a more meaningful way.

\begin{figure}
    \centering
    \includegraphics[width=\columnwidth]{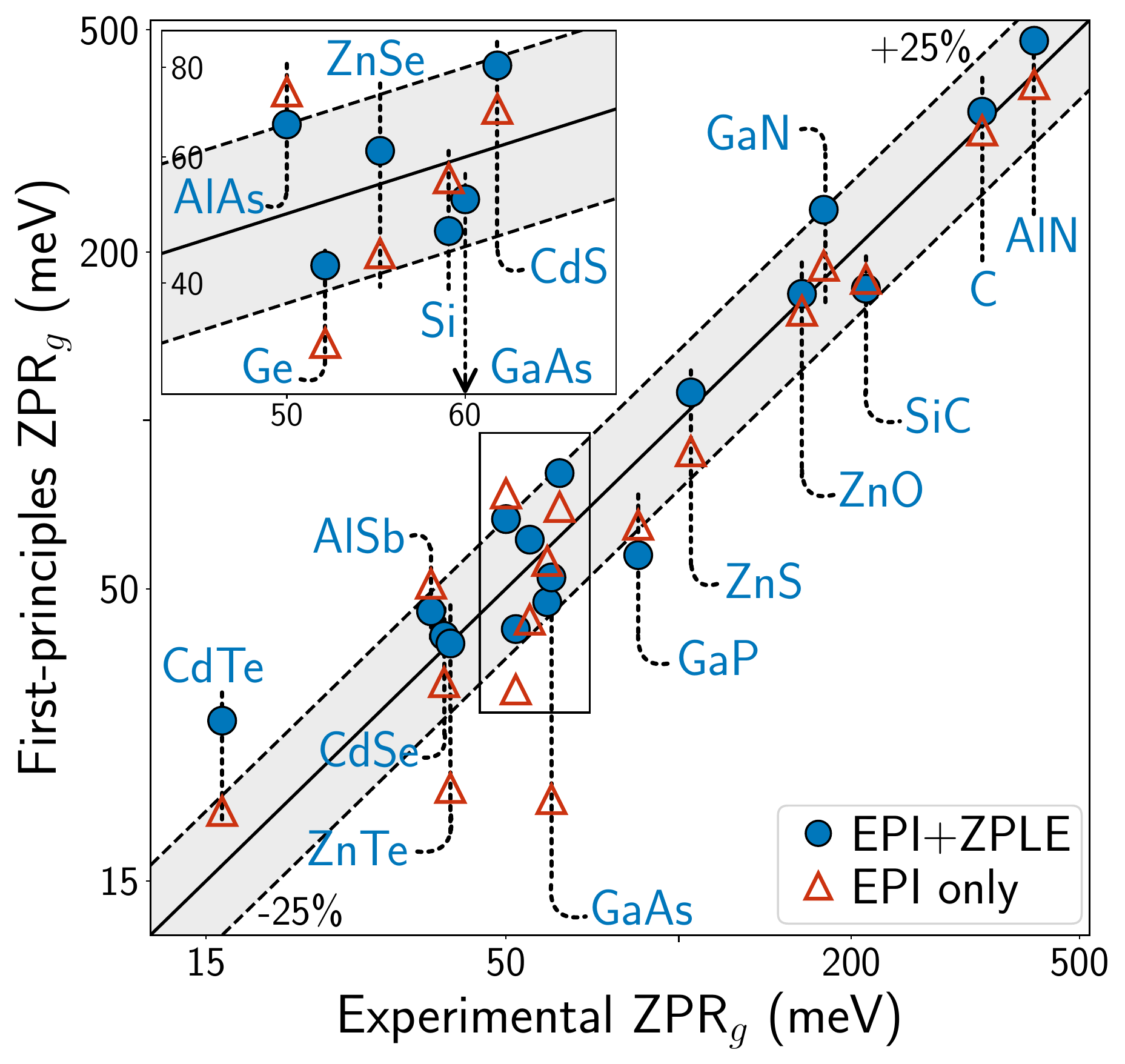}
    \caption{\textbf{Comparison of the predicted first-principles band gap ZPR to experimental values,} from EPI only (empty red triangles) and from the total EPI+ZPLE contributions (blue circles). The dashed lines and shaded gray area highlight the region where both values agree within 25\% of each other (see text). Note that the scales are logarithmic. When the ZPLE contribution is properly accounted for, more materials lie in this target region compared to when it is neglected. Detailed numerical values can be found in Table~S5 of the Supplemental Material~\cite{note_supplemental_material}.}
    \label{fig:zpr_vs_exp}
\end{figure}
In Fig.~\ref{fig:zpr_vs_exp}, we compare the accuracy of the predicted ZPR$_g$ to experimental values for 17 of our 22 materials, when either including (blue circles) or excluding (empty red triangles) the ZPLE contribution. All numerical values, including the reference experimental values, can be found in Table~S5 of the Supplemental Material~\cite{note_supplemental_material}. The dashed lines and shaded gray area highlight the region where both values agree within 25\% of each other (namely, when the smallest of the two values is 25\% smaller than the largest one, such that the gray area is symmetrical). As pointed out in the previous section for the fractional ZPLE, zero-point corrections are usually extracted from experimental data from extrapolation procedures, which yield an uncertainty which can be quite substantial. Hence, we believe that a 25\% agreement is acceptable.

From Fig.~\ref{fig:zpr_vs_exp}, we note that the effect of the ZPLE contribution on the experimental agreement is somewhat mitigated, as about half of the materials (9 materials) show an improved accuracy, while it lessens for the other half (8 materials). Nevertheless, the ZPLE contribution has an overall positive effect on the global agreement of the entire dataset, as more materials enter (Ge, GaAs, AlAs, ZnTe) than leave (CdTe, GaP) the shaded area when it is properly accounted for. For CdTe, the lack of spin-orbit coupling could explain the observed overestimation. Moreover, the relative agreement improves for most materials for which the ZPLE contribution is significant. Despite this, we stress that a decrease in accuracy for a given material does not justify purposely disregarding a real physical effect.

As the typical ZPLE correction is on the order of or smaller than the variation of the static lattice parameters between different exchange-correlation (XC) functionals, one can also wonder if the large ratios reported in Table~\ref{tab:ZPdata}, hence the experimental agreement discussed in Fig.~\ref{fig:zpr_vs_exp}, are simply an artifact of our choice of the PBE-GGA. See Ref.~\cite{He2014} for an investigation of the sensitivity of phonon frequencies to the XC functionals, including the effect of differing lattice constant. To verify the numerical stability of our results to the choice of functional, we computed the ZPR$_g^{\textrm{EPI}}$ and ZPR$_g^{\textrm{ZPLE}}$ contributions with the Perdew-Wang parametrization of the local density approximation (LDA)~\cite{perdew_accurate_1992} and the PBE approximation for solids (PBESOL)~\cite{perdew_restoring_2008} for five representative materials: the ZnX family (X=S, Se, Te), GaAs and MgO. These results are presented in Table~\ref{tab:ZPdata_functionals}. Despite their different respective static equilibrium lattice parameters, the fractional ZPLE remains on the same order for all three functionals. The slightly larger fractional ZPLE obtained from PBE can be understood in terms of the typical lattice parameter overestimation associated with this functional: as the material is slightly less rigid, the lattice response to the zero-point energy is marginally larger. Nevertheless, this difference has no noticeable effect on the resulting ZPR$_g^{\rm{FE}}$. We also find that the ratio R$^{\rm{ZPLE/EPI}}$ is not significantly affected by the choice of XC functional. It should be understood that these ratios are to be interpreted semi-quantitatively, as an indicator of the importance of the ZPR$_g^{\rm{ZPLE}}$ compared to the ZPR$_g^{\rm{EPI}}$ contribution. Hence, despite small variations (less than 5\% except for GaAs) between the different XC functionals, Table~\ref{tab:ZPdata_functionals} clearly establishes that the ratios presented in Table~\ref{tab:ZPdata} are robust within semilocal DFT. In the case of GaAs, it indicates that the ZPLE contribution to the total ZPR$_g$ is necessary to bridge the gap with experimental data, as it steadily contributes more than half of the total ZPR$_g$. A similar conclusion can also be drawn for ZnTe. 

However, whether the relative strength of the ZPLE contribution compared to EPI reported in Table~\ref{tab:ZPdata} maintains the same order when using higher-level XC functionals remains to be investigated. As a more exact treatment of electronic correlations has been shown to increase the strength of the ZPR$_g^{\rm{EPI}}$ in some materials~\cite{antonius_manybody_2014,li_electron-phonon_2019,li_unmasking_2021}, one could reasonably expect the ratios to decrease slightly. However, as Ref.~\cite{antonius_manybody_2014} reported a $10$~meV increase of the magnitude of the ZPR$_g^{\rm{EPI}}$ of GaAs when computing EPI with single-shot GW (from $-23$ to $-33$~meV), this suggests that electronic correlations alone are not sufficient to obtain experimental agreement for this material. The ZPLE contribution is thus most likely the missing ingredient.

Lastly, one fundamental assumption made throughout this work is that, following Eq.~(\ref{eq:sumzpr-full}), the total band gap ZPR can be approximated by the individual contributions of EPI and ZPLE. This supposes that the output of $\Sigma\kn^{\textrm{Fan}}$ (Eq.~(\ref{eq:sigmafan})) and $\Sigma\kn^{\textrm{DW}}$, 
which are computed at the static equilibrium lattice parameters, is not significantly affected by this small change of the lattice parameters, beyond what corresponds to the ZPLE effect on the band gap in the static DFT picture. In other words, it assumes that the QHA picture, which decorrelates both mechanisms and neglects the effect of anharmonicity in its description of EPI, is valid. We stress that, in principle, one should evaluate the temperature-dependent EPI self-energy at the temperature-dependent lattice parameter. While anharmonic effects can reasonably be expected to be small at absolute zero temperature, it remains to be investigated if computing the EPI self-energy at the ZPLE-renormalized lattice parameter instead of the bare lattice parameter, as was done for example in Ref.~\cite{zhang_temperature-dependent_2020}, would yield a total $\textrm{ZPR}_g$ that differs from the EPI-only picture by a similar ratio, even when the ZPLE contribution is large compared to EPI.

\subsection{Further discussion}\label{sec:discussion}
At this point, we would like to bring up the fact that zero-point corrections, whether on the lattice parameter, on the band gap energy or on any other physical quantity, cannot be obtained from a direct nor an indirect measurement. By essence, they represent the difference between a true quantum behavior of the lattice at $T=0$~K and a hypothetical lattice behavior, should the ions behave classically as $T\rightarrow0$~K and remain at rest. Hence, one cannot escape applying extrapolation procedures to measured data. Assessing the accuracy of the current numerical methods for computing zero-point corrections is thus no straightforward task.

In this work, we addressed the ZPLE question from a pure first-principles point of view, taking the FE minimization within the QHA as a reference value. We have not attempted to assess the accuracy of this framework with respect to experiments. Many studies have, however, partially addressed this question by investigating the temperature dependence of the lattice constants or the thermal expansion coefficients (see, for example, Refs.~\cite{malica_quasi-harmonic_2020, huang_efficient_2016, mounet_first-principles_2005, otero-de-la-roza_equations_2011}). As these works focus on reproducing the correct shape of the $T$-dependence, the numerical results are conveniently rescaled to the $T=0$~K measured values (such that, for example, the fractional ZPLE is null). In this picture, one can draw conclusions about the accuracy of numerical methods in different temperature regimes, but not about zero-point corrections.

Along the same line, another point brought to light by the results of Table~\ref{tab:zpgap-estimation} is the great sensitivity of Eq.~(\ref{eq:ZPLE-band_estimation}) to the actual value of of the fractional ZPLE, $\Delta a/a^0$. Naturally, they also depend on $B_0$ and $\partial E_g/\partial P$; still, it is quite unlikely that these well-studied quantities vary by more than a few percent (drastically) between experiments or numerical studies from static DFT. Hence, one must be particularly careful when using experimental data to obtain an experimental value of the fractional ZPLE. Again, we do not question the accuracy of published experimental data. We rather want to point out that not all available data are suitable for extracting experimental fractional ZPLE. On this subject, we also refer to the Supplementary Material of Ref.~\cite{miglio_predominance_2020} and to Ref.~\cite{monserrat_extracting_2014}, which discuss experimental evaluations of the total band gap ZPR from temperature-dependent band gap energies. As pointed out by our results, it should be understood that, unless the experimental temperature range extends well above the Debye temperature \emph{and} at sufficiently high temperatures, a fractional ZPLE extracted from a linearization and extrapolation procedure as described in Sec.~\ref{sec:res-estimate} should be thought of as a lower bound for the true ZPLE. While experimental results for temperature-dependent lattice parameters are abundant in the literature, datasets that fit both of these requirements are not. 
\section{Conclusion}
In this work, we investigated the zero-point lattice expansion within the quasi-harmonic approximation and its contribution to the total zero-point renormalization of the band gap using first-principles methodologies 
for a set of 22 benchmark semiconductors and insulators. We demonstrated that the formalism based on \gru parameters, which commonly neglects zero-point corrections, can efficiently predict the ZPLE both for isotropic and anisotropic materials, at a lower computational cost than the traditional Helmholtz free energy minimization method.

The \gru approach reproduces the results obtained from free energy minimization with a mean absolute relative error of 2.2\% for the linear lattice expansion, which is typically of the order of 0.1-0.3\% of the bare lattice parameter. The resulting ZPLE contribution to the band gap ZPR from both methodologies agrees within meV accuracy. Using our numerical results, we validated a posteriori an empirical expression that estimates the ZPLE contribution to the band gap ZPR from few simple parameters. The quality of this estimation is strongly sensitive to the temperature range chosen to extrapolate the fractional ZPLE, either from numerical or experimental data.

We finally assess the importance of the ZPLE contribution to the total band gap ZPR compared to the contribution of electron-phonon interaction. For heavier or more ionic materials, the inclusion of the ZPLE contribution typically modifies the predicted band gap ZPR by 20-80\%, including some materials containing light atoms. These proportions are shown to be stable within semilocal DFT. For GaAs, ZPLE is shown to be the missing ingredient to obtain agreement with the experimental band gap ZPR value. Both renormalization mechanisms must therefore be considered on an equal footing to deliver predictive theoretical ZPR values, and should be systematically incorporated in the different methodologies used to evaluate the band gap ZPR.

Overall, our findings bring to light that the decades-old \gru formalism can provide the community with a simple and accessible framework to incorporate zero-point lattice corrections in thermal expansion studies. Our results also indicate that a general understanding of the relative strength of the electron-phonon interaction and ZPLE contributions to the band gap ZPR in terms of simple material properties is still lacking. Investigating such a relationship from a broader perspective would further our fundamental understanding of zero-point quantum effects.
\begin{acknowledgments}
This research was financially supported by the Natural Sciences and Engineering Research Council of Canada (NSERC), under the Discovery Grants
program grant No.~RGPIN-2016-06666, 
and by the Fonds de la Recherche Scientifique (FRS-FNRS Belgium) through
the PdR Grant No.~T.0103.19 -~ALPS.
This research was enabled in part by support provided by Calcul Qu\'ebec (\url{www.calculquebec.ca}) and the Digital Research Alliance of Canada (\url{www.alliancecan.ca}). The operation of the supercomputers used for this research is funded by the
Canada Foundation for Innovation (CFI), the Minist\`ere de la Science, de l'\'Economie et de l'Innovation du Qu\'ebec (MESI), and the Fonds de recherche du Qu\'ebec – Nature et technologies (FRQ-NT). V.B.-C. acknowledges support by the NSERC Alexander
Graham Bell Canada Graduate Scholarship doctoral program, the FRQ-NT B2 Doctoral Scholarship and the Hydro-Québec Excellence Scholarship. V.B.-C, \'E.G. and M.C. are members of the Regroupement qu\'eb\'ecois sur les mat\'eriaux de pointe (RQMP).
\end{acknowledgments}
\clearpage
\section*{\Large SUPPLEMENTAL MATERIAL}
\setcounter{section}{0}
\setcounter{table}{0}
\setcounter{equation}{0}
\renewcommand{\thesection}{S\arabic{section}} 
\renewcommand{\thetable}{S\arabic{table}} 
\renewcommand{\theequation}{S\arabic{equation}} 
\section{\gru formalism at finite external pressure}\label{sec:SMgrun_pressure}
At finite external pressure $P$, one has to work with the Gibbs free energy (FE) of the combined electronic and lattice system which is approximated in the same fashion as described in the main text,
\begin{equation}\label{gibbsfefull}
    G(P, T) = \min_V \Big( E^{\text{e}}_{\textrm{stat}}(V) + F^{\text{vib}}(V,T) + PV \Big).
\end{equation}

For fixed pressure $P$ and temperature $T$, the equilibrium lattice configuration, $V_P\triangleq V_P(P,T)$, will verify
\begin{equation}\label{eq:equilibrium_condition}
   \left.\frac{\partial E^{\text{e}}_{\textrm{stat}}}{\partial V}
   \right|_{V=V_P}
   +
   \left.\frac{\partial F^{\text{vib}}}{\partial V}
   \right|_{V=V_P,T}
   +P
 = 0.
\end{equation}

In the following, we generalize the results of Sec.~\MakeUppercase{\romannumeral 2}~B.2 of the main text to finite external pressure for the volumic case only. Assuming that the internal coordinates can safely be approximated by those minimizing the Born-Oppenheimer static electronic energy at volume $V_P(P,T)$, the extension to anisotropic expansion is straightforward.

We first define the static equilibrium lattice configuration, $V_P^0$, which minimizes the electronic Gibbs FE, $G^{\text{e}}$, at fixed external pressure $P$,
\begin{equation}\label{eq:gibbse}
    G^{\text{e}}(P, T) = \min_V \Big( E^{\text{e}}_{\textrm{stat}}(V) + PV \Big).
\end{equation}
giving
\begin{equation}\label{eq:equilibrium_condition2}
   \left.\frac{\partial E^{\text{e}}_{\textrm{stat}}}{\partial V}
   \right|_{V=V_P^0}
   +P
 = 0.
\end{equation}
In all these equations, we neglected the entropic electronic contribution to the free energy as we are dealing with semiconductors and insulators.
From a computational point of view, $P$ is the external constraint applied on the static lattice during the relaxation process. 
It can be expressed as minus the slope of the static electronic energy evaluated at the static equilibrium volume $V_P^0$.

Coming back to the temperature-dependent case, in order to emphasize that $P$ is a target pressure, we will now denote it as $P_{\textrm{target}}$. 
The equilibrium condition on the total Gibbs FE (Eq.~(\ref{eq:equilibrium_condition})) yields the relationship
\begin{equation}\label{eq:pwithp}
    P_{\textrm{target}} = -
    \left.\frac{\partial E_{\textrm{stat}}^{\text{e}}}{\partial V}
    \right|_{V}
    -\left.\frac{\partial F^{\text{vib}}(V,T)}{\partial V}
    \right|_{V,T},
\end{equation}
where $P_{\textrm{target}}$ now contains the derivatives of both the electronic and phononic contributions. We can understand this equation as a search for the volume $V_P(P,T)$ for which the total Helmholtz FE at temperature $T$ has the same slope as in the static case.

Since $G^{\text{e}}$ is minimal at $V=V_P^0$, we Taylor-expand around this reference volume. The second-order expansion of the static electronic energy gives
\begin{equation}
\begin{split}
    E_{\textrm{stat}}^{\text{e}} (V) \approx& E_{\textrm{stat}}^{\text{e}}(V_P^0) + \left.\frac{\partial E_{\textrm{stat}}^{\text{e}}}{\partial V}
    \right|_{V_P^0}
    (V-V_P^0)\\ 
    &+ \frac{1}{2}\left.\frac{\partial^2 E_{\textrm{stat}}^{\text{e}}}{\partial V^2}
    \right|_{V_P^0} (V-V_P^0)^2,
    \end{split}
\end{equation}
while the expansion of $F^{\textrm{vib}}$ yields
\begin{equation}
   F^{vib} (V,T) \approx F^{vib}(V_P^0,T) + \left.\frac{\partial F^{\textrm{vib}}}{\partial V}\right|_{V=V_P^0,T} \left(V-V_P^0\right),
\end{equation}
where the partial derivative is identical to the second line of Eq.~(12) of the main text, with $V^0$ replaced by the new reference volume $V_P^0$. 
Substituting these expansions in Eq.~(\ref{eq:pwithp}), we obtain
\begin{equation}\label{eq:pderiv}
\begin{split}
    P_{\textrm{target}} \approx& -\frac{\partial}{\partial V}\bigg(-P_{\textrm{target}}(V-V_P^0)\\ &+ \frac{1}{2} \left.\frac{\partial^2 E_{\textrm{stat}}^{\text{e}}}{\partial V^2}\right|_{V_P^0} (V-V_P^0)^2 - \left. \frac{\partial F^{\text{vib}}}{\partial V}\right|_{V=V_P^0,T}(V-V_P^0)\bigg)\\
    \approx& \;P_{\textrm{target}} - \left.\frac{\partial^2 E^{\text{e}}}{\partial V^2}\right|_{V_P^0} (V-V_P^0) - \left.\frac{\partial F^{\text{vib}}}{\partial V}\right|_{V=V_P^0,T}.
    \end{split}
\end{equation}
The second derivative in the first term on the right-hand side can be recast in terms of the static bulk modulus evaluated at constant pressure $P_{\textrm{target}}$:
\begin{equation}\label{eq:d2estrain}
\begin{split}
    \left.\frac{\partial^2 E_{\textrm{stat}}^{\text{e}}}{\partial V^2}\right|_{V_P^0} &= \left.-\frac{\partial P}{\partial V}\right|_{V_P^0}\\
    &= \frac{B_0(P_{\textrm{target}})}{V_P^0}.
    \end{split}
\end{equation}
Substituting Eq.~(\ref{eq:d2estrain}) in Eq.~(\ref{eq:pderiv}) and cancelling out the $P_{\textrm{target}}$ terms,
we recover an expression analogous to Eq.~(12) of the main text, in which the reference static configuration used to evaluate the equilibrium volume, bulk modulus, and phonon frequencies already captures the effect of finite external pressure on the lattice. The resulting expression for the ZPLE defined in the main text can therefore be applied to finite external pressure by using a strained reference configuration.
\section{Generalized Grüneisen ZPLE limit for cubic and axial crystals}\label{sec:SMsymm}
In the following, we show that Eq.~(27) of the main text reduces to the well-known literature results for cubic and axial crystals and provide explicit expressions for the Gr\"uneisen ZPLE (Eq.~(28) of the main text).
In the case of cubic symmetry, the only non-zero coefficients in the compliance tensor are $s_{11}=s_{22}=s_{33}$, $s_{12}=s_{13}=s_{23}$ and $s_{44}=s_{55}=s_{66}$. Furthermore, the infinitesimal strains $\epsilon_i,\;{i=4,5,6}$ are null, as the crystal symmetry is preserved. There is only one independent lattice constant, hence $\epsilon_1=\epsilon_2=\epsilon_3$ and $\gamma\qv^1=\gamma\qv^2=\gamma\qv^3$. Note that $\epsilon_1=\Delta a/a^0$ by definition of the stress tensor components. Eq.~(27) of the main text thus reduces to~\cite{grimvall_thermophysical_1986}
\begin{equation}\label{eq:sm-alpha-for-cubic}
    \begin{split}
        \alpha(T) &= \frac{1}{V^0}\sum\qv \left(s_{11}\gamma\qv^1 + s_{12}\gamma\qv^2 + s_{13}\gamma\qv^3\right) \wqv c\qv^V(T)\\
        &= \frac{1}{V^0} \left(s_{11}+2s_{12}\right)\sum\qv \gamma\qv^1\wqv c\qv^V(T)\\
        &= \frac{1}{3B_0V^0 }\sum\qv \gamma\qv^1\wqv c\qv^V(T),
    \end{split}
\end{equation}
since $s_{11}+2s_{12} = (3B_0)^{-1}$. In this expression, $\gamma\qv^1$ refers to the derivative taken by straining only \emph{one} of the three lattice parameters, the other two remaining at their fixed static equilibrium value. One could recast this term to compute the derivative with respect to a simultaneous change of lenght of \emph{all three} lattice vectors at the cost of a factor 1/3 :
\begin{equation}
\begin{split}
    \gamma\qv^1 &= -\frac{a_i}{\wqv}\left.\frac{\partial \wqv}{\partial a_i}\right|_{a_{j\neq i} = a^0} \\ &= -\frac{1}{3}\frac{a^0}{\wqv}\left(\frac{\partial \wqv}{\partial  a}\right)\\
    &= \frac{1}{3}\gamma^a\qv
\end{split}
\end{equation}
The ZPLE expression for cubic symmetry (Eq.~(28) of the main text) then reads
\begin{equation}
\begin{split}
    \Delta a(T=0) &= \frac{a^0}{3B_0V^0}\sum\qv\gamma\qv^1\frac{\wqv}{2}\\
    &= \frac{a^0}{9B_0V^0}\sum\qv\gamma\qv^a\frac{\wqv}{2}
    \end{split}
\end{equation}

One can also show that the anisotropic case contains the volumic result, Eq.~(15) of the main text. First, from Eq.~(34) of the main text, we know that $\beta (T) = 3\alpha(T)$ for cubic materials. Hence, 
\begin{equation}\label{eq:sm-deltav-cubic}
\begin{split}
    \frac{ \Delta V (T)}{V^0} &= 3\frac{\Delta a(T)}{a^0 },\\
    &= \frac{1}{3B_0V^0}\sum\qv\gamma\qv^a\wqv\left(n\qv (T) + \frac{1}{2}\right),
\end{split}
\end{equation}
where we reintroduced the temperature dependence using the relations shown in  Eq.~(25) to~(27) of the main text. We now find the relation between the linear mode \gru parameters $\gamma\qv^a$ and their volumic counterpart,
\begin{equation}
   \begin{split}
      \gamma\qv^a &= -\frac{a^0}{\wqv}\left.\frac{\partial \wqv}{\partial a}\right|_{a^0}\\
      &= -\frac{a^0}{\wqv}\left.\frac{\partial \wqv}{\partial V}\frac{\partial V}{\partial a}\right|_{a^0}\\
      &= -\frac{a^0}{\wqv}(3(a^0)^2)\left.\frac{\partial \wqv}{\partial V}\right|_{V^0}\\
      &= 3\left(-\frac{V^0}{\wqv}\left.\frac{\partial \wqv}{\partial V}\right|_{V^0}\right)\\
      &= 3\gamma\qv^V.
   \end{split}
\end{equation}
Substituting this result in Eq.~(\ref{eq:sm-deltav-cubic}), we recover the volumic expression, Eq.~(15) of the main text,
\begin{equation}
   \frac{ \Delta V (T)}{V^0}= \frac{1}{B_0V^0}\sum\qv\gamma\qv^V\wqv\left(n\qv (T) + \frac{1}{2}\right). 
\end{equation}

For axial symmetry (i.e. hexagonal, trigonal and tetragonal lattices), there are only two non-zero components of the mode Gr\"uneisen tensor,  $\gamma\qv^1=\gamma\qv^2$ and $\gamma\qv^3$. Hence, the relevant independent compliance coefficients are $s_{11}=s_{22}$, $s_{33}$, $s_{12}$ and $s_{13}=s_{23}$. We obtain
\begin{equation}
    \begin{split}
        \alpha^1 (T) &= \frac{1}{V^0}\sum\qv \left( (s_{11}+s_{12})\gamma\qv^1 + s_{13}\gamma\qv^3\right)\wqv c\qv(T) \\
        &= \frac{1}{V^0}\sum\qv \left(\frac{1}{2}(s_{11}+s_{12})\gamma\qv^a + s_{13}\gamma\qv^c\right)\wqv c\qv(T),\\
        \alpha^3(T) &= \frac{1}{V^0}\sum\qv \left( 2s_{13}\gamma\qv^1 + s_{33}\gamma\qv^3\right)\wqv c\qv(T)\\
        &= \frac{1}{V^0}\sum\qv \left( s_{13}\gamma\qv^a + s_{33}\gamma\qv^c\right)\wqv c\qv(T),
    \end{split}
\end{equation}
which are the expressions tabulated in the literature~\cite{munn_gruneisen_1968}. The ZPLE thus takes the form
\begin{equation}
    \begin{split}
                \Delta a(T=0) &= \frac{a^0}{V^0}\sum\qv \left( \frac{1}{2}(s_{11}+s_{12})\gamma\qv^a + s_{13}\gamma\qv^c\right)\frac{\wqv}{2}, \\
        \Delta c(T=0)
        &= \frac{c^0}{V^0}\sum\qv \left( s_{13}\gamma\qv^a + s_{33}\gamma\qv^c\right)\frac{\wqv}{2}.
    \end{split}
\end{equation}

\section{Calculation parameters for the different materials}\label{sec:SMcalcdetails}

Table~\ref{tab:Calcdata} specifies the Materials Project ID number, space group, cut-off energy for the plane-wave basis set for the GS, phonon and elastic constants calculations, \kpoint sampling of the BZ, as well as the \qpoint sampling used for the ZPLE and EPI calculations, for the 22 materials investigated for this work. For elastic constant calculations, the aforementioned cutoff energies were used in conjunction with a cutoff energy smearing of 0.5 hartree, to prevent discontinuities in the total energy curve when computing shear strains. The larger increase of the cutoff energy for AlN can be attributed to older pseudopotentials. For more complete details about the EPI calculation for ZnO, see the METHODS section and Supplementary Material of Ref.~\cite{miglio_predominance_2020}.
\begin{table*}[h]
    \centering
    \caption{\textbf{Calculations parameters for the twenty-two materials in our set.} From left to right, material ID on the Materials Project, cut-off energy for the plane-wave basis set for GS/phonons (E$_{\textrm{cut}}$) and elastic constants (E$_{\textrm{cut}}, c_{ij}$) calculations, \kpoint sampling for the BZ, and \qpoint sampling for ZPLE and $\textrm{ZPR}_g^{\textrm{EPI}}$ ({$N\times N\times N$} Monkhorst-Pack grids). All \qpoint grids are $\Gamma$-centered. The elastic constants cutoff energy was used in conjunction with a cutoff energy smearing of 0.5 hartree. See Sec.~\MakeUppercase{\romannumeral 3}\,B of the main text for more details regarding the \qpoint sampling for GaAs.}
    \label{tab:Calcdata}
     \setlength\extrarowheight{2pt}
        \begin{tabularx}{\textwidth}{p{17mm} p{17mm} c c c c c c}
    \hline\hline
      \multirow{2}{*}{Material}  & \multirow{2}{*}{MP-ID} & \multirow{2}{*}{Space group} & \multirow{2}{*}{E$_{\textrm{cut}}$ (Ha)} & \multirow{2}{*}{E$_{\textrm{cut}}, c_{ij}$ (Ha)} & \multirow{2}{*}{\kpoint sampling} & {\qpoint sampling} &  {\qpoint sampling}\\
      & & & & &  & {(ZPLE)} & {(EPI)~\cite{miglio_predominance_2020}}\\[2pt]
      \hline
      C & 66 &{Fd$\bar{3}$m [227]}&35 &{45}& {$6\times6\times6$ ($\times$4 shifts)}& {$8\times 8\times 8$} & {$125\times 125\times 125$~\cite{ponce_temperature_2015}}\\
      Si & 149 &{Fd$\bar{3}$m [227]} &25 &{25} & {$6\times6\times6$ ($\times$4 shifts)}& {$8\times 8\times 8$} & {$100\times 100\times 100$}\\
        Ge & 32 & {Fd$\bar{3}$m [227]} &40 &{40}& {$6\times6\times6$ ($\times$4 shifts)} & {$8\times 8\times 8$} & {$48\times 48\times 48$}\\
        SiC &  8062 & {F$\bar{4}3$m [216]}&35 &{45}& {$6\times6\times6$ ($\times$4 shifts)}& {$8\times 8\times 8$}& {$48\times 48\times 48$}\\
        \hline
        BN & 1639 &{F$\bar{4}3$m [216]} &35 &{40}& {$8\times 8\times 8$  ($\times$4 shifts)}& {$8\times 8\times 8$} & {$100\times 100\times 100$~\cite{ponce_temperature_2015}}\\
        BAs & 10044 & {F$\bar{4}3$m [216]} &40&{45} & {$6\times6\times6$ ($\times$4 shifts)}& {$8\times 8\times 8$}& {$48\times 48\times 48$}\\  
        AlP & 1550 &{F$\bar{4}3$m [216]}&25 &{30}& {$8\times 8\times 8$ ($\times$4 shifts)}& {$8\times 8\times 8$}& {$48\times 48\times 48$}\\
        AlAs & 2172 &{F$\bar{4}3$m [216]} &40 &{40}& {$6\times6\times6$ ($\times$4 shifts)}& {$8\times 8\times 8$} & {$48\times 48\times 48$}\\
        AlSb & 2624 &{F$\bar{4}3$m [216]} &40 &{45}& {$6\times6\times6$ ($\times$4 shifts)}& {$8\times 8\times 8$}& {$48\times 48\times 48$}\\
        GaN & 830 &{F$\bar{4}3$m [216]} &40 &{45}& {$6\times6\times6$ ($\times$4 shifts)}& {$8\times 8\times 8$}& {$48\times 48\times 48$}\\
        GaP & 2690 & {F$\bar{4}3$m [216]}&40 &{40}& {$6\times6\times6$ ($\times$4 shifts)}& {$8\times 8\times 8$} & {$48\times 48\times 48$}\\
        GaAs & 2534 &{F$\bar{4}3$m [216]} &40 &{40}& {$6\times6\times6$ ($\times$4 shifts)}& {$8\times 8\times 8$} & {$64\times 64\times 64$}\\
        \hline
        ZnS & 10695 & {F$\bar{4}3$m [216]}&40 &{40}& {$6\times6\times6$ ($\times$4 shifts)}& {$8\times 8\times 8$}& {$48\times 48\times 48$}\\
        ZnSe & 1190 & {F$\bar{4}3$m [216]}&40 &{40}& {$6\times6\times6$ ($\times$4 shifts)}& {$8\times 8\times 8$} & {$48\times 48\times 48$}\\
        ZnTe & 2176 &{F$\bar{4}3$m [216]} &40 &{40}& {$6\times6\times6$ ($\times$4 shifts)}& {$8\times 8\times 8$}& {$48\times 48\times 48$}\\
        CdS &  2469 &{F$\bar{4}3$m [216]} &45 &{50}& {$6\times6\times6$ ($\times$4 shifts)}& {$8\times 8\times 8$}& {$48\times 48\times 48$}\\
        CdSe & 2691 & {F$\bar{4}3$m [216]}&50 &{50}& {$6\times6\times6$ ($\times$4 shifts)}& {$8\times 8\times 8$} & {$48\times 48\times 48$}\\
        CdTe & 406 & {F$\bar{4}3$m [216]}&50 &{55}& {$6\times6\times6$ ($\times$4 shifts)}& {$8\times 8\times 8$} & {$48\times 48\times 48$}\\
        MgO & 1265 & {Fm$\bar{3}$m [225]}&50 &{50}& {$8\times 8\times 8$ ($\times$4 shifts)}& {$8\times 8\times 8$}& {$96\times 96\times 96$}\\
        \hline
        AlN & 661 & {P6$_3$mc [186]}&35 &{60}& {$6\times6\times6$}  & {$8\times 8\times 8$}& {$34\times 34\times 34$~\cite{ponce_temperature_2015}}\\
        GaN & 804 &{P6$_3$mc [186]} &40 &{45}& {$8\times 8\times 8$} & {$8\times 8\times 8$}& {$64\times 64\times 64$}\\
        ZnO & 2133 & {\mbox{P6$_3$mc} [186]} &50 &{50}& {$6\times6\times6$}& {$8\times 8\times 8$} & {$48\times 48\times 48$}\\
          \hline\hline
    \end{tabularx}
\end{table*}
\section{Additional numerical tables}
Tables~\ref{tab:a0datab0} and~\ref{tab:a0c0datab0} reproduce the ZPLE results of Tables~\MakeUppercase{\romannumeral 1} and~\MakeUppercase{\romannumeral 2} of the main text, making explicit the absolute relative error with respect to FE minimization results for each material.

Table~\ref{tab:deltaa_on_a} displays the numerical values of the fractional ZPLE ($\Delta a\,(T=0)/a^0$ and $\Delta c\,(T=0)/c^0$) used in Eq.~(37) of the main text to obtain the different estimations of the $\textrm{ZPR}_g^{\textrm{ZPLE}}$ reported in Table~\MakeUppercase{\romannumeral 6} of the main text. Note that, since we use $\Delta E_g^{\textrm{Grun 1\%}}$ as a reference, $\Delta a^{\textrm{Grun 1\%}}/a^0$ and $\Delta a^{\textrm{data}}/a^0$ are by construction identical. For $\Delta a^{700-1000\textrm{K}}/a^0$ and $\Delta a^{T>2\theta_D}/a^0$, we used the ZP-renormalized lattice constants, $a(T=0)$, as the reference configuration to define the temperature-dependent strain:
\begin{equation}
    \frac{\Delta a\,(T)}{a^0}\rightarrow \frac{a(T) - a(T=0)}{a(T=0)}
\end{equation}
This definition is slightly different from Eq.~(17) of the main text. Since the purpose of this analysis is to evaluate the reliability of Eq.~(37) of the main text when applied to experimental data, we chose to treat our theoretical data as if it were experimental results.
As $a\,(T=0) = a^0 + \Delta a\,(T=0)$, this induces an error if the order of ($\Delta a\,(T=0)/a^0)^2$, which will not affect our conclusions. Note finally that such an error is naturally present in the evaluation of $\Delta a^{\textrm{exp}}/a^0$.\\
The reported data in the rightmost column was obtained from the following procedures. For Ge, Si and diamond, we average the $\Delta a(T=0)/a^0 $ data reported in Sec.~\MakeUppercase{\romannumeral 3}.B.2 of Ref.~\cite{cardona_isotope_2005} from 3 different techniques, our definition of $\Delta a(T=0)/a^0$ differing by a sign change from theirs. For AlN and GaN, we use the experimental data displayed in Figs.~(1) and~(3) of Ref.~\cite{wang_thermal_1997} from references therein. For GaN, this  value of $\Delta a\,(T=0)/a^0$ is averaged with the one extracted from Fig.~(1) of Ref.~\cite{roder_temperature_2005}, from markers labelled \enquote{this work} and \enquote{Ref.~10}, to obtain the $\textrm{ZPR}_g$ estimation reported in the main text. For ZnO, we extract $\Delta a\,(T=0)/a^0$ by combining high-temperature data from Ref.~\cite{iwanaga_anisotropic_2000} and low-temperature data from Ref.~\cite{reeber_lattice_1970}. Note that this method may be biased by a small lattice parameter mismatch between the two combined datasets, as they have been obtained from different samples and measuring devices.

Table~\ref{tab:sm-zp-expdata} finally reports the experimental values shown in Fig.~5 of the main text and compares them to our numerical results. When available, we favored values extracted from isotopic substitution data, which is less sensitive to the extrapolation procedure as the high-temperature asymptote method described in Sec.~\MakeUppercase{\romannumeral 4}D of the main text (see Supplementary Note~1 of the Supplementary Information of Ref.~\cite{miglio_predominance_2020} for a more detailed analysis of the available experimental ZPR$_g$ data). For the latter method, we rely on the works of P\"assler~\cite{passler_dispersion_2002,passler_semiempirical_2003}. For AlN, we set our reference value as the average of the reported values from those two articles, resp. $-350$~meV and $-483$~meV.


\begin{table}[]
    \centering
    \caption{\textbf{Zero-point lattice expansion for cubic materials: absolute relative error (ARE) of the \gru approach with respect to FE minimization.} $a^0$ is the theoretical lattice parameter in the static lattice approximation, except for Ge and GaP (see Sec.~\MakeUppercase{\romannumeral 3}\,A of the main text for details). The variation of $a^0$ induced by the zero-point motion of the ions, $\Delta a^{\textrm{FE}} (T=0)$, is obtained by minimizing the Helmholtz free energy, while $\Delta a^{\textrm{Grun}} (T=0)$ is obtained from the \gru parameters, using Eq.~(28) of the main text.} 
    \label{tab:a0datab0}
      \setlength\extrarowheight{2pt}
  \sisetup{
table-number-alignment = center ,
table-column-width = 1.3cm,
}        
\begin{tabularx}{\columnwidth}{@{} p{14mm} C{1.3cm}
S[table-format = 1.4] S[table-format = 1.4] S[table-format = 1.2] S[table-format = 1.2] @{}}
\hline\hline
\multirow{2}{*}{Material}   &\multirow{2}{*}{\shortstack{$\Delta a^{\textrm{FE}}$ \\ (Bohr)}} &  \multicolumn{2}{c}{{$\Delta a^{\textrm{Grun}}$ (Bohr)}} &\multicolumn{2}{c}{{ARE vs FE (\%)}} \\
\cmidrule(lr){3-4}\cmidrule(lr){5-6}
& & {1\%} & {0.5\%} & {1\%} & {0.5\%} \\[2pt]
       \hline
      C-dia &  0.0253 & 0.0250 & 0.0248 & 1.19 & 1.98\\
      Si-dia &  0.0174 & 0.0172 &0.0172 & 1.15 & 1.15\\
      Ge-dia & 0.0102 & 0.0112 & 0.0112 & 9.80 & 9.80\\
      SiC-zb & 0.0227 & 0.0226&0.0223 & 0.44 & 1.76\\
      \hline
      BN-zb & 0.0285 & 0.0271 &0.0270 & 5.01 & 5.26\\
      BAs & 0.0238 & 0.0237 & 0.0236 & 0.42 & 0.84\\
      AlP-zb &  0.0180 & 0.0178 &0.0178  & 1.11 & 1.11\\
      AlAs-zb &  0.0166 & 0.0162& 0.0163 & 2.41 & 1.81\\
      AlSb-zb & 0.0163 & 0.0162 & 0.0161 & 0.61 & 1.23\\
      GaN-zb &0.0209 &0.0208 &0.0208 & 0.48 & 0.48\\
      GaP-zb &  0.0152 & 0.0156&0.0156 & 2.63 & 2.63 \\
      GaAs-zb &  0.0157 & 0.0146 & 0.0145 & 7.01& 7.64\\
      \hline 
      ZnS-zb &  0.0194 &0.0193& 0.0192 & 0.52 & 1.03\\
      ZnSe-zb &  0.0160 & 0.0157& 0.0155 & 1.88 & 3.13\\
      ZnTe-zb & 0.0154 & 0.0153& 0.0153 & 0.65 & 0.65\\
      CdS-zb &0.0177 & 0.0170& 0.0162 & 3.95 & 8.47\\
      CdSe-zb & 0.0136 & 0.0133 & 0.0135 & 2.21 & 0.74\\
       CdTe-zb &  0.0129 & 0.0126 & 0.0125 & 2.33 & 3.10\\
    MgO-rs & 0.0333 & 0.0317& 0.0318 & 5.09 & 4.79\\
\hline\hline
    \end{tabularx}
\end{table}


\begin{table}[]
    \centering
    \caption{\textbf{Zero-point lattice expansion for wurtzite materials: absolute relative error (ARE) of the \gru approach with respect to FE minimization.} $a^0$ and $c^0$ are the theoretical lattice parameter in the static lattice approximation. The variation of $a^0$ and $c^0$ induced by the zero-point motion of the ions, resp. $\Delta a^{\textrm{FE}} (T=0)$ and $\Delta c^{\textrm{FE}} (T=0)$, are obtained by minimizing the Helmholtz FE, while $\Delta a^{\textrm{Grun}} (T=0)$ and $\Delta c^{\textrm{Grun}} (T=0)$ are calculated from the \gru parameters, using Eq.~
(28) of the main text.} 
    \label{tab:a0c0datab0}
      \setlength\extrarowheight{2pt}
 \sisetup{
table-number-alignment = center ,
table-column-width = 1.3cm,
}
    \begin{tabularx}{\columnwidth}{@{} p{14mm} 
    C{1.3cm}
    S[table-format = 1.4] S[table-format = 1.4] S[table-format = 1.2] S[table-format = 1.2] @{}}
    \hline\hline
    \multirow{2}{*}{Material} & \multirow{2}{*}{\shortstack{$\Delta a^{\textrm{FE}}$ \\ (Bohr)}} & \multicolumn{2}{c}{{$\Delta a^{\textrm{Grun}}$ (Bohr)}} &\multicolumn{2}{c}{{ARE vs FE (\%)} }\\
    \cmidrule(lr){3-4}\cmidrule(lr){5-6}
    & & {1\%} & {0.5\%}& {1\%} & {0.5\%}\\[2pt]
    \hline
    AlN & 0.0170 & 0.0163& 0.0162 & 4.12 & 4.71 \\
    GaN & 0.0151 & 0.0151 &0.0151  &0.00 & 0.00\\
    ZnO & 0.0165 & 0.0165 &0.0167&  0.00 & 1.21\\
    \hline
    \multirow{2}{*}{Material}  &\multirow{2}{*}{\shortstack{$\Delta c^{\textrm{FE}}$ \\ (Bohr)}}&  \multicolumn{2}{c}{{$\Delta c^{\textrm{Grun}}$ (Bohr)}} &\multicolumn{2}{c}{{ARE vs FE (\%)}} 
       \\
     \cmidrule(lr){3-4}\cmidrule(lr){5-6}
     & & {1\%} & {0.5\%} & {1\%} & {0.5\%} \\[2pt]
       \hline
    AlN  & 0.0241 & 0.0242 &0.0242
        & 0.42 & 0.42\\
    GaN & 0.0232 & 0.0228 &0.0229
        & 1.73 & 1.29\\
    ZnO & 0.0213 & 0.0213 &0.0207 
        & 0.00 & 2.82\\
          \hline\hline
          \end{tabularx}
\end{table}
\begin{table}[]
    \caption{\textbf{Numerical values of $\bm{\Delta a(T=0)/a^0}$ and $\bm{\Delta c(T=0)/c^0}$ used to obtain estimations of the ZPLE band gap ZPR in Table~\MakeUppercase{\romannumeral 6} of the main text.} At linear order, we can express the $\Delta V(T=0)/V^0$ term entering Eq.~(37) of the main text
    as $3\Delta a(T=0)/a^0$ for cubic materials, and as $2\Delta a(T=0)/a^0+ \Delta c(T=0)/c^0$ for wurtzites.  The different methodologies used to evaluate the fractional ZPLE are described in Sec.~\MakeUppercase{\romannumeral 4}\,C.2
    of the main text. Note that, since the \gru 1\% results as a reference value in Table~\MakeUppercase{\romannumeral 6},
 $\Delta a^{\textrm{Grun 1\%}}(T=0)/a^0$ = $\Delta a^{\textrm{data}}(T=0)/a^0$.}
    \label{tab:deltaa_on_a}
\sisetup{
table-number-alignment = center,
table-format = 1.2,
}
     \setlength\extrarowheight{2pt}
    \begin{tabularx}{\columnwidth}{@{} p{13mm}  S[table-column-width = 1.3cm] S[table-column-width = 1.35cm] S[table-column-width = 1.80cm] S[table-column-width = 2.1cm]
    @{}}
    \hline\hline
       \multirow{3}{*}{Material}  & {\multirow{2}{*}{$\mathlarger{\frac{\Delta a}{a^0}^{\textrm{data}}}$}} & {\multirow{2}{*}{$\mathlarger{\frac{\Delta a}{a^0}^{T>2\theta_D}}$}} & {\multirow{2}{*}{$\mathlarger{\frac{\Delta a}{a^0}^{700-1000\textrm{K}}}$}}&  {\multirow{2}{*}{$\mathlarger{\frac{\Delta a}{a^0}^{\textrm{exp}}}$}}\\
       & & & & \\
       &{(10$^{-3}$)} &{(10$^{-3}$)} &{(10$^{-3}$)} &{(10$^{-3}$)} \\[2pt]
       \hline
    C-dia   & 3.70& 3.17 &1.67    & {3.9, 3.7, 3.9}~\cite{cardona_isotope_2005}\\
    Si-dia   &1.66 & 1.43 & 1.23   & {1.7, 1.9, 2.0}~\cite{cardona_isotope_2005}\\
    Ge-dia   &  1.06 & 0.90 & 0.89  & {1.3, 1.2, 1.8}~\cite{cardona_isotope_2005}\\
    SiC-zb   & 2.73 & 2.39&1.65 & 2.71~\cite{reeber_thermal_1995}\\
    \hline
    BN-zb   & 4.02 & 3.36 &1.99\\
    BAs    & 2.60 & 2.22 &1.75  & 1.30~\cite{kang_basic_2019}\\
    AlP-zb   & 1.71 & 1.48 &1.30\\
    AlAs-zb   & 1.50 & 1.27& 1.22 & \\
    AlSb-zb  &  1.38 & 1.11& 1.15\\
    GaN-zb    & 2.42 & 1.99&1.68\\
    GaP-zb     & 1.52 & 1.28 & 1.21  &1.73~\cite{reeber_thermal_1995}\\
    GaAs-zb   &  1.34 & 1.17 &1.16 &\\
    \hline
    ZnS-zb     & 1.88 & 1.59&1.58\\
    ZnSe-zb    & 1.45 & 1.29& 1.28\\
    ZnTe-zb     &  1.31 &1.09 &1.18\\
    CdS-zb     & 1.52 & 1.19&1.29\\
    CdSe-zb    & 1.14 & 0.94 &1.02\\
    CdTe-zb   &  1.01 & 0.84& 0.92\\
    MgO-rs   & 3.95 & 3.61&3.07 & {4.01~\cite{dubrovinsky_thermal_1997}, 3.77~\cite{fiquet_high-temperature_1999}}\\
    \hline
    AlN-w  &  2.81 & 2.44& 1.80& 2.64~\cite{wang_thermal_1997}\\
    {GaN-w}    & 2.48 &1.91 & 1.72 & {1.13~\cite{wang_thermal_1997}, 2.50~\cite{roder_temperature_2005}}\\
    ZnO-w    & 2.67 & 2.15  &2.00 & {4.23~\cite{iwanaga_anisotropic_2000, reeber_lattice_1970}}\\
    \hline\hline
      \multirow{3}{*}{Material}  & {\multirow{2}{*}{$\mathlarger{\frac{\Delta c}{c^0}^{\textrm{data}}}$}} & {\multirow{2}{*}{$\mathlarger{\frac{\Delta c}{c^0}^{T>2\theta_D}}$}} & {\multirow{2}{*}{$\mathlarger{\frac{\Delta c}{c^0}^{700-1000\textrm{K}}}$}}&  {\multirow{2}{*}{$\mathlarger{\frac{\Delta c}{c^0}^{\textrm{exp}}}$}}\\
       & & & & \\
       &{(10$^{-3}$)} &{(10$^{-3}$)} &{(10$^{-3}$)} &{(10$^{-3}$)} \\[2pt]
       \hline
       AlN-w  &  2.62&2.26 & 1.64& 2.71~\cite{wang_thermal_1997}\\
       GaN-w    & 2.30 & 1.78& 1.60& {1.41~\cite{wang_thermal_1997}, 1.91~\cite{roder_temperature_2005}}\\
        ZnO-w   & 2.14 & 1.73  & 1.62& {1.19~\cite{iwanaga_anisotropic_2000, reeber_lattice_1970}}\\
        \hline\hline
    \end{tabularx}
\end{table}
\begin{table}[]
    \centering
    \caption{\textbf{Comparison of the predicted band gap ZPR to experimental values} from the EPI contribution only and from the combined EPI+ZPLE contributions (see Table~\MakeUppercase{\romannumeral 8} of the main text). The unitless ratio $R^{\rm{tot}}$ is defined as $R^{\rm{tot}}=\textrm{ZPR}_g^{\rm{tot}}/\textrm{ZPR}_g^{\rm{exp}}$, and reported as \mbox{$R^{\rm{tot}}=1/(\textrm{ZPR}_g^{\rm{exp}}/\textrm{ZPR}_g^{\rm{tot}})$} when \mbox{$|\textrm{ZPR}_g^{\rm{tot}}|>|\textrm{ZPR}_g^{\rm{exp}}|$}, in order to emphasize the symmetric boundaries of the shaded gray area in Fig.~5 of the main text  
    (see Sec.~\MakeUppercase{\romannumeral 4}~C~3 of the main text for details). The ratio $R^{\rm{EPI}}$ is defined in a similar manner, using $\textrm{ZPR}_g^{\rm{EPI}}$. $^\dagger$The reference value for AlN is the average value of the reported values from Ref.~\cite{passler_dispersion_2002} and~\cite{passler_semiempirical_2003} (resp. $-350$~meV and $-483$~meV).}
    \label{tab:sm-zp-expdata}
\sisetup{
table-format = 4.2 ,
table-number-alignment = center ,
table-column-width = 1.33cm ,
}
     \setlength\extrarowheight{2pt}
    \begin{tabularx}{\columnwidth}{p{13mm} 
    S[table-format=5.0] 
    S[table-format=5.0] 
    S[table-number-alignment = left] S S}
    \hline\hline
       \multirow{2}{*}{Material}   & {ZPR$_g^{\rm{EPI}}$} & {ZPR$_g^{\rm{tot}}$} & {ZPR$_g^{\rm{exp}}$} & {$R^{\rm{EPI}}$} &  {$R^{\rm{tot}}$}\\
        & {(meV)} & {(meV)} & {(meV)} & & \\[2pt]
       \hline
      C-dia & -330 & -357  & {$-338$~\cite{collins_indirect_1990}} & {0.98} & {1/0.95} \\
      Si-dia  & -56 & -47 & {~~$-59$~\cite{karaiskaj_photoluminescence_2002}} & {0.95} & {0.80}\\
        Ge-dia & -33 & -42 & {~~$-52$~\cite{parks_electronic_1994}} & {0.63} & {0.81}\\
        SiC-zb & -179 & -172 & {$-212$~\cite{passler_semiempirical_2003}} & {0.84} & {0.81}\\
        \hline
        AlAs-zb & -74 & -67 & {~~$-50$~\cite{passler_dispersion_2002}} & {1/0.68} & {1/0.75}\\
        GaP-zb & -65& -57 & {~~$-85$~\cite{passler_dispersion_2002}} & {0.76} & {0.68}\\
        GaAs-zb & -21 & -52 &  {~~$-60$~\cite{passler_dispersion_2002}} & {0.35} & {0.87}\\
        \hline
        ZnS-zb & -88 & -112 & {$-105$~\cite{manjon_effect_2005}} & {0.84} & {1/0.94}\\
        ZnSe-zb & -44 & -61 & {~~$-55$~\cite{passler_dispersion_2002}} & {0.80} & {1/0.90}\\
        ZnTe-zb & -22 & -40 & {~~$-40$~\cite{passler_dispersion_2002}} & {0.55} & {1.00}\\
        CdS-zb & -70 & -80 & {~~$-62$~\cite{zhang_isotope_1998}} & {1/0.89} & {1/0.77}\\
        CdSe-zb & -34 & -41 & {~~$-39$~\cite{passler_semiempirical_2003}} & {0.87} & {1/0.95}\\
        CdTe-zb & -20 & -29 & {~~$-16$~\cite{passler_dispersion_2002}} & {1/0.80} & {1/0.55}\\
        \hline
        AlN-w & -399 & -478 & {\!\!\!\!\!\!\!\!$-417^\dagger$} & {0.96} & {1/0.87}\\
        GaN-w & -189 & -238  & {$-180$~\cite{passler_dispersion_2002}} & {1/0.95} & {1/0.76}\\
        ZnO-w & -157 & -168 & {$-164$~\cite{manjon_effect_2003}} & {0.96} & {1/0.97}\\
          \hline\hline
    \end{tabularx}
\end{table}
\FloatBarrier
%

\end{document}